\pdfoutput=1
\documentclass[11pt]{article}
\usepackage[utf8]{inputenc}
\usepackage[margin=1in]{geometry}

\usepackage{amsmath}
\usepackage{amssymb}
\usepackage{amsthm}
\usepackage{color}
\usepackage{amsfonts}
\usepackage{bbm}
\usepackage[makeroom]{cancel}

\usepackage[toc,page]{appendix}

\usepackage{pifont} 

\usepackage{graphicx}
\usepackage{float}
\usepackage{caption}
\usepackage{verbatim}

\newtheorem{theorem}{Theorem}[section]
\newtheorem{lemma}[theorem]{Lemma}
\newtheorem{proposition}[theorem]{Proposition}
\newtheorem{corollary}[theorem]{Corollary}
\newtheorem{definition}[theorem]{Definition}

\title{Optimal Solutions for Adaptive Search Problems with Entropy Objectives}
\author{Huanyu Ding and David Casta{\~n}{\'o}n%
\thanks{This work was supported by NSF award CNS-1330008.  The authors are with the Dept. of Electrical \& Comp. Eng., Boston University.
        {\footnotesize hyding@bu.edu, dac@bu.edu}}%
}

\date{}

\begin{document}

\maketitle

\begin{abstract}
The problem of searching for an unknown object occurs in important applications ranging from security, medicine and defense.  Sensors with the capability to process information rapidly require adaptive algorithms to control their search in response to noisy observations.   In this paper, we discuss classes of dynamic, adaptive search problems, and formulate the resulting sensor control problems as stochastic control problems with imperfect information, based on previous work on noisy search problems.   The structure of these problems, with objective functions related to information entropy, allows for a complete characterization of the optimal strategies and the optimal cost for the resulting finite-horizon stochastic control problems.   We study the problem where an individual sensor is capable of searching over multiple sub-regions in a time, and provide a constructive algorithm for determining optimal policies in real time based on convex optimization.   We also study the problem in which there are multiple sensors, each of which is only capable of detecting over one sub-region in a time, jointly searching for an object.  Whereas this can be viewed as a special case of our multi-region results, we show that the computation of optimal policies can be decoupled into single-sensor individual scalar convex optimization problems, and provide simple symmetry conditions where the solutions can be determined analytically. We also consider the case where individual sensors can select the accuracy of their sensing modes with different costs, and derive optimal strategies for these problems in terms of the solutions of scalar convex optimization problems.  We illustrate our results with experiments using multiple sensors searching for a single object. 

\end{abstract}

\section{Introduction}
The proliferation of intelligent sensors  in diverse applications from building security, defense, transportation and medicine has created a need for automated processing and deduction of sensor information.  An important problem in these sensor systems is the detection and localization of objects of interest.  Intelligent sensors are able to control the nature of information collected by changing their field of view and their sensing parameters; ideally, they should do so adaptively, exploiting what has been learned from previous observations, to improve the accuracy in detection and localization.  In this paper, we focus on the problem of developing adaptive search policies for a stationary object in a compact domain, with sensors that provide noisy information regarding the presence of the object in the field of view. 

The field of search theory has a long history, dating back to its early application for locating submarines and objects at sea in the 1940's \cite{Koopman1946,KoopmanBook}.  The search problem was formulated as an optimal allocation of search effort to look for a single stationary object with a single imperfect sensor \cite{StoneBook,Stone1977,Benkoski1991}.  The sensor detects the presence of the object, with a simple sensor error model, a probability of missed detection (but not a corresponding probability of a false alarm).  The resulting search strategies were  open-loop search plans, which continued until an output of ``detected'' was returned.  The limitations of  this approach were that the sensor measurements could produce no false alarms, and that resulting search strategies were non-adaptive.  Extensions of search theory to more complex error models that require adaptive feedback strategies have been developed in some restricted contexts \cite{Castanon1995} where a single sensor can observe one of many possible discrete locations at each time.   

In the presence of complex noise models, the adaptive search problem can be viewed as a problem of sensor management, looking for optimal controlled sensing policies \cite{DavidsBook}.  There are many  applications of sensor management techniques that develop adaptive strategies for different sensing problems, including function estimation \cite{acLearn1}, image acquisition \cite{faceDetect}, object classification  \cite{castanon2005stochastic,recedeHorizon,infoBased,markovObj}, object tracking \cite{williams_fisher_willsky07,kreucher_hero_kastella_morelande07,kreucher2005sensor}.  These problems can be formulated as instances of stochastic control problems \cite{dpbook} and sequential experiment design \cite{seqExp1, seqExp2, seqExp3}.    However, exact solution of these stochastic dynamic programming problems is computationally prohibitive, so most of these adaptive techniques use heuristics and approximations such as model-predictive control to obtain strategies with manageable computation complexity.   

Our formulation to the adaptive search problem is based on the approach proposed by Jedynak et al. \cite{20qs}.  In their work, 
Jedynak et al. \cite{20qs} considered the problem of localizing a stationary object in an Euclidean space by using a single sensor that asks a sequence of yes/no questions, each of which asks whether the object is located in a region specified by the sensor.   We refer to the questions as \textit{sensing modes} in our paper, and the decisions are made on selecting the sensing modes.   The sensor observes a Boolean value corresponding to whether or not the object is localized in the inquired region, but this yes/no value is corrupted by noise at the output of the sensor.  We refer to such a sensor  as a \textit{Boolean sensor}.  Jedynak et al. formulated the optimal Boolean sensing problem to optimize the posterior differential entropy of the object location after a fixed finite number of measurements, and showed the existence of optimal strategies as well as explicit constructions for the optimal adaptive strategies for several variations of this problem \cite{20qs}.  

The problem studied in \cite{20qs} has its roots in information theory known as the R\'enyi-Ulam game \cite{CoverBook}.  Horstein  \cite{horstein1963sequential} developed  a probabilistic bisection scheme for the noisy version of this problem in the context of sequential decoding.  Burnashev and Zigangirov  \cite{BZ_Algorithm} developed an algorithm for the case where the possible query locations are discrete and showed asymptotic decay in the probability of location error.    Nowak  \cite{nowak2008generalized} proposed a generalized binary search algorithm to search over a discrete location space.  The R\'enyi-Ulam game with adversarial errors was studied in \cite{dhagat1992playing, spencer1992ulam}.  

Recently, the work in \cite{20qs} has been generalized in several directions.  Sznitman et al.  \cite{microscopy} considered the case where the sensors can choose different types of observations with different costs, with application to problems in electron microscopy.   Tsiligkaridis et al. \cite{hero2014collaborative} considered the problem of multiple Boolean sensors performing collaborative search, where each sensor observes a noisy measurement of the Boolean indicator that the object is contained in the observed subset.  They developed characterizations of optimal strategies for the multi-sensor case.   Focusing on the case where each sensor has a binary symmetric error model, they provided explicit analytic solutions for the optimal multistage adaptive sensing policies. In subsequent work \cite{hero2015decentralized}, they extended their strategies to the decentralized case where sensors do not know each other's error models, but exchange local estimates of the conditional density of the object location, and provide a consensus algorithm where neighboring sensors exchange these local estimates to arrive at a common estimate of the conditional density of the object location.  

In this paper, we generalize the Boolean single-sensor search problem of \cite{20qs} to a pair of different search problems:  First, we consider the  \textit{multi-region} single-sensor search problem where the sensor can partition the object space into multiple regions and inquire as to which region the object is located in.  Second, we consider extensions of the Boolean multi-sensor search problem where there are multiple sensors working simultaneously as a team, similar to the problem considered in  \cite{hero2014collaborative}, but using more general error models.  The second problem can be viewed as a special case of the first problem with some added structure.   We adopt a Bayes formulation similar to that in \cite{20qs}, using general sensor error models, with the goal of reducing the final entropy of the conditional probability density of the object location after a known fixed number of observations.   We pose the first problem as a stochastic control problem, and derive a complete characterization of optimal adaptive strategies.   We also provide a constructive algorithm for computing the optimal strategies based on convex optimization, and show that the optimal strategies are independent of the problem horizon.  We further derive a lower bound on the performance of the minimum mean-square error estimator.   

For the Boolean multi-sensor search problem, we show the equivalence of this problem to our multiregion search problem, thereby establishing a characterization of optimal policies and the optimal cost.  We further show that the optimal sensing strategies can be obtained in terms of the solution of decoupled scalar convex optimization problems, by showing that the optimal joint policies have a special factorization structure that can be obtained from the solution of individual Boolean single sensor problems.  We also show the equivalence in expected performance between a system where multiple sensors collect information simultaneously, and one where sensors collect information sequentially among sensors, with information from each sensor shared instantaneously so it can be incorporated into the choice of other sensors' actions.  We describe a generalized symmetry condition  for non-binary error models that enables the analytic solution of the joint sensing problem, and provide a constructive solution for generating the optimal adaptive sensing strategies. 

As a further extension, we consider the scenario where each Boolean sensor is also allowed to choose among error models for their observations at different costs, extending the results of \cite{microscopy} to the multi-sensor case.   We derive the optimal policies for this problem of costly Boolean multi-sensor search.  We provide explicit solutions for the optimal value function, and show that the optimal strategies can again be computed in terms of the solution of single-sensor problems.  We provide experiments with two and three sensor simulations that illustrate the performance of our sensing strategies.  

The results of this paper are a significant extension of our previous results reported in \cite{DingCastanon15}, which focused mostly on the single sensor multiregion search problem.  Even for that problem, our exposition presents a more rigorous treatment of the optimality conditions with full proofs, a new symmetry condition that enables analytic solution for determining optimal strategies, and a lower bound on the performance of mean square estimation error.   The results in this paper illustrate how the structure of entropy-based objectives can lead to complete characterizations of optimal adaptive sensing strategies, along with practical algorithms for computation of such strategies. 

The paper is structured as follows: In Section \ref{sec:multianswer}, we study the multi-region single-sensor search problem.  We describe the problem formulation, and derive the optimal policies for this model.  Based on the optimal cost,  we provide a the lower bound on the covariance of the minimum mean-square error estimator of the object's unknown location.   In Section \ref{sec:multisensor}, we study the Boolean multi-sensor search problem with general sensor error models.   We describe the model formulation, and develop the optimal solution of the model, similar to the results of \cite{hero2014collaborative}.  We show that, for general discrete error models, the optimal policies can be determined through the solution of decoupled single sensor problems, and provide a simple construction for those problems.    In Section  \ref{subsec:precision_mode}, we study the multi-sensor search problem where sensors can control both the choice of sensing mode as well as the precision mode of that search, in terms of a choice of error models, given a cost of selecting the observation mode, generalizing the work of \cite{microscopy}.  We develop a simple computational algorithm for selecting optimal policies for sensing mode and precision mode selection.   Section \ref{sec:simulation_multisensor} contains simulation results using both two sensors and three sensors that illustrate the performance of our approaches.   Section \ref{sec:conclusion} contains conclusions and directions for future work.   The Appendix contains the proofs of the major results.

\section{Multi-Region Single-Sensor Search with Noise}
\label{sec:multianswer}

\subsection{The Multi-Region Single-Sensor Model}
\label{subsec:multianswer_model}
Consider the problem of localizing a stationary point object  whose position is denoted by $X$,  a continuous-valued random vector in a compact subset $\mathcal{X}$ of  $\mathbb{R}^d$ ($d \leq 3$ for our purposes) with prior probability distribution that is absolutely continuous with respect to Lebesgue measure, with density $p_0(x)$.  We assume this initial density has finite differential entropy.  We have a single sensor, which can collect measurements of the object location. As in \cite{20qs, hero2014collaborative} and previous approaches to search theory \cite{StoneBook}, we avoid modeling explicit sensor locations and activities, and instead model sensor measurements as aggregate efforts over a domain of interest.  In our formulation, a sensing mode is a  partition of the domain $\mathcal{X}$ into $K \ge 2$ disjoint Lebesgue measurable regions $ \{ A^{(i)} \}$, with each assigned the distinct integer label $i$ in $\{1, \ldots, K\}$.  A sensing mode will result in observed measurement values for the sensor.  We assume that the sensors collect measurements in discrete stages by choosing its sensing mode at each stage.

In the absence of measurement noise, the value of the measurement would correspond to identifying which region contains the object $X$. That is, 
\begin{align*}
 Z = \sum_{i=1}^K i \mathbbm{1}_{\{X \in A^{(i)}\}}
\end{align*}

In our formulation, the sensor measurements include noise. The measurement obtained by the sensor, $Y$, will be a random variable that can be either discrete or continuous valued.  For the rest of this paper, we assume that  $Y$ is discrete-valued, with values in a discrete set $\mathcal{Y}$.  Our results extend in straightforward manner to the case where $Y$ takes values in a continuous space.  We assume measurements can be collected at each stage $n$, with the measurement noise is defined by the conditional probability distribution of $Y_n$  given the value $Z_n$:
\begin{align*}
P(Y_n=y | Z_n = k) = f_k(y), \quad k = 1,\ldots, K
\end{align*}
We assume the measurements $Y_n, n = 1, \ldots, N$ are conditionally independent given the object location $X$ and sensing modes $A_n, n = 1, \dots, N$.  

Our goal is to obtain $N$ measurements sequentially to improve our knowledge of the object location $X$.  Let $A_n$ denote the sensing mode used for the measurement at stage $n$: this partitions 
 $\mathcal{X}$ into $K$ disjoint regions $\{ A_n^{(1)}, \cdots, A_n^{(K)} \}$, and let $Y_n$ denote the measurement obtained under that mode.  The information history collected by the sensor after the $n$-th measurement is denoted as
\begin{align*}
D_n = \{ A_1,Y_1,\cdots,A_n,Y_n\}
\end{align*}
 
Let $p_n(x) = p(x|D_n)$ denote the posterior density of $X$ given the history $D_n$.  We denote this quantity as the information state at stage $n$.  The evolution of this information state across stages is derived using Bayesian reasoning as follows:  Assume we know $p_n(x)$, and we obtain a measurement $Y_{n+1}=y$ given sensing mode $A_{n+1}$.  Then,
\begin{align}
p_{n+1}(x)
& = p_n(x) \cdot \frac{P(Y_{n+1}=y|A_{n+1},X=x)}{\int_{\mathcal{X}} p_n(\sigma) \cdot P(Y_{n+1}=y|A_{n+1},X=\sigma) d\sigma} \nonumber \\
& = p_n(x) \cdot \frac{\sum_{k=1}^K f_k(y) \mathbbm{1}_{\{x\in A^{(k)}_{n+1} \}}}{\int_{\mathcal{X}} p_n(\sigma)  \sum_{k=1}^K f_k(y) \mathbbm{1}_{\{\sigma\in A^{(k)}_{n+1} \}}            d\sigma} \label{Bayesrule} 
\end{align}

The above evolution can be viewed as a stochastic dynamical system for the information state $p_{n+1}(x)$, where the evolution depends on the finite-valued random ``disturbance'' $y$, with conditional probability density $\eta(y) = {\int_{\mathcal{X}} p_n(\sigma)  \sum_{k=1}^K f_k(y) \mathbbm{1}_{\{\sigma\in A^{(k)}_{n+1} \}}            d\sigma}$ that depends on the current information state $p_n(x)$ and the control action $A_{n+1}$.  As long as $\eta(y) > 0$, the resulting information state $p_{n+1}$ is well-defined, and will represent a probability density on $\mathcal{X}$.    For $\eta(y) = 0$, we  arbitrarily define $p_{n+1}(x) = p_n(x)$. 

A useful quantity in our development is $P(Z_{n+1} = k | p_n, A_{n+1}) \equiv u_{n+1}^{(k)}$, computed as
\begin{equation}
 u_{n+1}^{(k)} = \int_{A_{n+1}^{(k)}} p_n(\sigma) d\sigma \ \geq 0
\label{stats}
\end{equation}
We refer to $\boldsymbol{u}_{n+1} = (u_{n+1}^{(1)},\ldots,u_{n+1}^{(K)})$ as the \textit{operating point} at stage $n+1$.  Note that  $\sum_{k=1}^K u^{(k)}_{n+1} = 1$ because $A_{n+1}$ is a Lebesgue-measurable partition of $\mathcal{X}$.  With this notation, the denominator in Bayes' rule \eqref{Bayesrule} becomes 
$\eta(y) = \sum_{k=1}^K u^{(k)}_{n+1} f_k(y)$.

Let $\Gamma^K(\mathcal{X})$ denote the set of all partitions of the domain $\mathcal{X}$ into $K$ measurable subsets.  Let $S_n$ denote the space of probability densities $p_n(x)$ over $\mathcal{X}$, corresponding to distributions that are absolutely continuous with respect to Lebesgue measure.  We define an adaptive sensing {policy} $\pi = (\pi_1, \pi_2, \cdots, \pi_N)$ to be a sequence of functions $\pi_n: S_{n-1} \rightarrow \Gamma^K(\mathcal{X})$, which will map the information state  $p_{n-1}(\cdot)$ into the sensing mode used at stage $n$.  Let $\Pi$ denote the space of all adaptive sensing policies.

To finalize the problem formulation, we define the objective function.  We will evaluate the quality of our knowledge of $X$ after collecting information $D_n$ by its  \textit{posterior differential entropy} $H(p_n)$  defined as
\begin{align*}
H(p_n) = -\int_\mathcal{X} p_n(x) \log_2 p_n(x) dx
\end{align*}

Our goal is to minimize $H(p_N)$ --- the posterior differential entropy after  $N$ measurements. The problem of interest is to choose the adaptive policy  $\pi$ to minimize $H(p_N)$:
\begin{align}
\label{FinalGoal}
\underset{\pi \in \Pi}{\inf}\;E[H(p_N) | p_0]
\end{align}

Note that our objective is a nonlinear functional of the information state, unlike the standard models for partially observed Markov decision processes \cite{dpbook} where the final objective is a linear functional of the final information state.  

The above dynamic decision problem can be viewed as a perfectly observed Markov decision problem with infinite-dimensional state space $S_n$, stochastic dynamics with discrete-valued disturbances \eqref{Bayesrule}, and terminal cost objective \eqref{FinalGoal}.  The admissible control space  $\Gamma^K(\mathcal{X})$ for each information state $p_n(\cdot)$ has none of the typical topological structure (e.g. a Borel space or a metric space) assumed in most dynamic programming results.  Still, our problem satisfies the structure for stochastic optimal control with countable disturbances described in Chapter 3 of \cite{Ber_Shreve}.  
We define the optimal value function $V(p_n,n)$ at the stage step $n$ to be:

\begin{align}
V(p_n,n)=\inf_{(\pi_{n+1}, \ldots, \pi_{N})} E[H(p_N)|p_n]  \nonumber
\end{align}

The optimal value function has to satisfy the Bellman equation \cite{Ber_Shreve}:
\begin{align}
\label{BellmanEquation}
V(p_n,n) = \inf_{A_{n+1}} E_{Y_{n+1}}[V(p_{n+1},n+1)|A_{n+1},p_n]
\end{align}
Furthermore, if a policy $\pi^*$ satisfies 
$$  E_{Y_{n+1}}[V(p_{n+1},n+1)|\pi^{*}_{n+1}(p_n), p_n] = V(p_n,n)$$
for all $p_n$, then the policy is optimal.

\subsection{Optimal Policies for Multi-Region Single-Sensor Search}
\label{subsec:optimal_policy_multianswer}

To derive the optimal policy, we consider the reduction in expected posterior differential entropy $H(p_{n}) - E[H(p_{n+1})|A_{n+1},p_n]$ that results from a sensing mode $A_{n+1}$ based on information state $p_n(x)$.  The following proposition summarizes our result:

\begin{proposition}
\label{prop:OneStepEntropyUpdate_multianswer}
The expected reduction in posterior differential entropy from a sensing mode $A_{n+1}$ is given in terms of the operating points $\boldsymbol{u}_{n+1} = (u^{(1)}_{n+1}, \ldots, u^{(K)}_{n+1})$ in \eqref{stats}, as
\begin{align*}
H(p_n) - E_{Y_{n+1}}[H(p_{n+1})|A_{n+1},p_n] = \varphi(\boldsymbol{u}_{n+1})
\end{align*}
where
\begin{small}
\begin{align*}
\varphi(\boldsymbol{u}_{n+1})
& = \mathcal{H}(\sum_{k=1}^K f_k(y) u^{(k)}_{n+1}) - \sum_{k=1}^K u^{(k)}_{n+1} \mathcal{H}(f_k(y))
\end{align*}
\end{small}
\noindent where $\mathcal{H}$ is the standard Shannon entropy for discrete-valued distributions.  
\end{proposition}

The proof is shown in the Appendix.   One way of interpreting  $\varphi(\boldsymbol{u}_{n+1})$ is to consider $\boldsymbol{u}_{n+1}$ as a probability distribution for the values of a discrete-valued random variable $Z_{n+1}$, with $P(Z_{n+1} = j) = u^{(j)}_{n+1}$.  Then,  
$$\varphi(\boldsymbol{u}_{n+1}) = I(Y_{n+1};Z_{n+1})$$
where the mutual information for two discrete-valued random variables is defined in terms of the Shannon entropy as 
$$ I(Y;Z) = \mathcal{H}(Y) - \mathcal{H}(Y|Z)$$
This is readily established as
\begin{small}
\begin{align*}
 I(Y;Z) &= -\sum_{y \in \mathcal{Y}} \sum_{z \in \{1, \ldots, K\}} P(y|z)P(z) \log \Bigr [\sum_{z \in \{1, \ldots, K\}} P(y|z)P(z)   \Bigl ]+  \sum_{z \in \{1, \ldots, K\}}P(z) \sum_{y \in \mathcal{Y}} P(y|z) \log P(y|z) \\
&= -\sum_{y \in \mathcal{Y}} \sum_{k \in \{1, \ldots, K\}} u^{(k)}f_k(y) \log \Bigr [\sum_{k \in \{1, \ldots, K\}} u^{(k)}f_k(y)   \Bigl ] + \sum_{k \in \{1, \ldots, K\}}u^{(k)} \sum_{y \in \mathcal{Y}} f_k(y) \log f_k(y) \\
&= \mathcal{H}(\sum_{k=1}^K f_k(y) u^{(k)}) - \sum_{k=1}^K u^{(k)} \mathcal{H}(f_k(y))
\end{align*}
\end{small}

Note that 
$\varphi(\boldsymbol{u})=\varphi(u^{(1)},\cdots,u^{(K)})$ as defined in Proposition \ref{prop:OneStepEntropyUpdate_multianswer} is strictly concave over the simplex
$\sum_{k=1}^K u^{(k)} = 1$, for $u^{(k)} \geq 0$, $k = 1,\cdots,K$. This follows from the strict concavity of the Shannon entropy $\mathcal{H}(f)$.  Thus, it has a unique maximum value achieved at a unique point $\boldsymbol{u}^* = (u^{(1)*},\cdots,u^{(K)*})$.  Any partition $A_{n+1}$ for which the statistics in \eqref{stats} are equal to $\boldsymbol{u}^*$ achieves the maximal differential entropy reduction at stage $n+1$.  Note that the optimal operating point $\boldsymbol{u}^*$ does not depend on the posterior density $p_n(x)$ or the partition $A_{n+1}$.  

Next, we show that, for any operating point $\boldsymbol{u}^*$ and information state $p_n(x)$, there exists a sensing mode with partition $A_{n+1}$ for which $\boldsymbol{u}(A_{n+1},p_n) = \boldsymbol{u}^*$.  
Let $d$ denote the dimension of the Euclidean space containing  $\mathcal{X}$, and let $e$ denote the $d$-dimensional vector of all 1s.  Since $p_n(x)$ corresponds to a distribution that is absolutely continuous, the cumulative distribution function $P_n(x) = \int_{-\infty}^x \cdots \int_{-\infty}^x  p_n(x')  dx'$  is continuous, and monotone nondecreasing on the diagonal $x = \alpha e$, starting at 0 for $\alpha <= - C$, and increasing to 1 for $\alpha >=C$ for some $C$ because of the compactness of $\mathcal{X}$.  Hence, for any $u^{(1)*}$, we can find a value $a_1$ so that $P(a_1 e) = u^{(1)*}$, and we can set $A_{n+1}^{(1)} = \{x \le a_1 e\} \cap \mathcal{X}$, where the inequality is interpreted element wise.   Similarly, for any $u^{(2)*}$ such that $u^{(1)*} + u^{(2)*} \le 1$, we can find $a_2 \ge a_1$ such that $P(a_2 e) - P(a_1 e) = u^{(2)*}$, and set $A_{n+1}^{(2)} = \{ a_1 e < x \le a_2 e\} \cap \mathcal{X}$.  We continue this construction to obtain the final  $a_K = C$, because $\sum_{k=1}^K u^{(k)*} = 1$.  The final partition $A_{n+1}$ so constructed satisfies $\boldsymbol{u}(A_{n+1},p_n) = \boldsymbol{u}^*$.  Note that there are many other partitions that would also satisfy this equality, which implies that the optimal partition is not unique.  

What remains is to show the optimal solution to the multistage adaptive policy optimization problem \eqref{FinalGoal} can be constructed in terms of the above adaptive sensing policy. 

\begin{proposition}[Optimal Policies for Single-Sensor Multi-Region Search]
\label{prop:greedyIsOptThm_multianswer}
Let $(u^{(1)*},\cdots,u^{(K)*})$ $= \arg\max_{\boldsymbol{u}=(u^{(1)},\cdots,u^{(K)})} \varphi(\boldsymbol{u})$ for $\varphi(\boldsymbol{u})$ as defined in Proposition \ref{prop:OneStepEntropyUpdate_multianswer}.  For each stage $n$, select a sensing mode $A_n$ that satisfies $\boldsymbol{u}(A_{n+1},p_n) = \boldsymbol{u}^*$.  Then, this adaptive set of policies is optimal for problem \eqref{FinalGoal}.  Furthermore, the optimal value function is given by
\begin{align} \label{eq:value_fun_multianswer}
V(p_n,n)=H(p_n) - (N-n)\varphi^*
\end{align}
where the constant $\varphi^* = \varphi(u^{(1)*},\cdots,u^{(K)*})$.
\end{proposition}

The proof is included in the Appendix.  We note at this point that the optimal single stage entropy reduction $\varphi^*$ is equal to the information-theoretic channel capacity $C$ of a memoryless communication channel with input the discrete variables $Z$ and output the observations $Y$: both quantities are defined by the same optimization problem. 

An important property of the above solution is that the optimal feedback strategy does not depend on the length of the planning horizon $N$.  Thus, the resulting strategies are optimal for any duration of the planning horizon, resulting in search algorithms that are optimal no matter when the search terminates.

The above results exploit several special structures of our adaptive control problem, as discussed below:
\begin{itemize}
\item The object location must have a prior distribution over a continuous region that is absolutely continuous with respect to Lebesgue measure.  This leads to conditional cumulative probability distributions that are continuous, and enable us to construct strategies that satisfy the optimality conditions.  This would not be the case if the potential object locations had distributions that were not absolutely continuous. 
 
\item The differential entropy objective function allows for separability of the contribution of new information from past information, a critical step in the development of optimality conditions.  Replacing the objective by similar functions such as R\'enyi entropy or other similar divergence measures requires additional conditions to guarantee concavity as well as existence of minimizing strategies. 

\item The measurement error models do not depend on the size of the regions used in the partitions at each stage, and depend on $X$ only through the indicator that $X$ is in particular regions.   
\end{itemize}

There are special cases where the optimal solution is known explicitly.  One such case is when the measurement error model satisfies a special symmetry condition.  The error model from $Z$ to $Y$ is modeled as a noisy discrete memoryless channel.   This quasi-symmetry condition requires that the set of outputs $\mathcal{Y}$ can be partitioned into subsets $\mathcal{Y}^{(m)}$ such that, for each subset,  the sub-transition probability matrices $P(y | z) $ for $y \in \mathcal{Y}^{(m)}, z \in \{1, \ldots, K\}$ satisfy the property that the each row is a permutation of every other row, and each column sums up the same subset-dependent constant.  When this channel has the property of quasi-symmetry \cite{Chen00}, or otherwise satisfies the property of symmetry as defined in \cite{Gallager}, the optimal operating point satisfies  $(u^{(1)*},\cdots,u^{(K)*}) = (1/K, \ldots, 1/K)$ (\cite{Gallager}, Thm 4.5.2).

\subsection{Mean-Square Error Lower Bound on Performance of Multi-Region Single-Sensor Search}
\label{subsec:mse_lower_bound}
From Proposition \ref{prop:greedyIsOptThm_multianswer}, the maximal expected posterior entropy reduction is $n \varphi^*$ after $n$ sensing stages are completed, where $\varphi^*$ is the same as defined in Proposition \ref{prop:greedyIsOptThm_multianswer}.   This allows us to give a lower bound on the performance of the minimum mean-square error estimator, similar to the results in \cite{hero2014collaborative}:

\begin{proposition}[Mean-Square Error Lower Bound]
\label{prop:mse_lower_bound}
Assume $H(p_0)$ is finite.  Then, the minimum mean-square error estimator at stage $n$ $ \hat{X}_n = \int_{\mathcal{X}} x p_n(x) dx $ under any admissible policy has the following mean-square error lower bound:
\begin{align*}
E[||X - \hat X_n||_2^2] \geq \frac{d \sqrt[d]{C_0}}{2 \pi e} e^{-\frac{2n\varphi^*}{d}}
\end{align*}
where $d$ is the dimension of the object space and $C_0 = e^{2H(p_0)}$, and  $\varphi^*$ is defined in Proposition \ref{prop:greedyIsOptThm_multianswer}.  
\end{proposition}

This lower bound decays exponentially with the number of stages, at a rate that is proportional to the maximal one-stage expected entropy reduction $\varphi^*$.

\section{Boolean Multi-Sensor Search with Noise}
\label{sec:multisensor}

\subsection{The Boolean Multi-Sensor Model}
\label{subsec:multisensor_model}

As a special case of our previous results, we consider a problem where there are  $M$ ( $M \geq 2$) Boolean sensors; each sensor can select a single region of observation, which is a  Lebesgue measurable subset of $\mathcal{X}$, and receive a noisy answer as to whether the object $X$ is in the region.   The sensors  simultaneously collect measurements at each stage, and coordinate their sensing modes to develop adaptive sensing strategies.   This search model is similar to the joint sensing model studied in \cite{hero2014collaborative}, although our emphasis is on non-binary, non-symmetric error models whereas most of  \cite{hero2014collaborative} focuses on binary symmetric error models.  We assume that each sensor collects discrete-valued measurements, each of which takes values in a discrete set $\mathcal{Y}$.  
A decision at stage $n$ consists of selecting sensing modes $(A^{(1)}_n, \ldots, A^{(M)}_n)$ for all $M$ sensors in a batch, where $A^{(1)}_n, \ldots, A^{(M)}_n \subset \mathcal{X}$.  

Given the sensing mode for sensor $m$, the error-free measured value corresponds to whether or not the queried subset $A^{(m)}_n$ contains the object  $X$ as before:
\begin{align*}
Z^{(m)}_n = \mathbbm{1}_{\{X \in A^{(m)}_n\}}, \quad \mbox{for sensor $m$ at stage $n$}
\end{align*}

As before, we assume that the noisy measurements are discrete-valued, taking values in a discrete alphabet $\mathcal{Y}$.  The statistical distribution of the collected noisy measurement for each sensor $m$ is
\begin{align*}
P(Y^{(m)}_n = y^{(m)} | Z^{(m)}_n = i_m) = f_{i_m}^{(m)} (y^{(m)}), \quad y^{(m)}\in\mathcal{Y}, \  i_m\in\{0,1\}, \  \  \mbox{for sensor $m$ at stage $n$}
\end{align*}

\noindent
We assume that the noisy measurements $Y^{(1)}_n, \ldots, Y^{(M)}_n$ are conditionally independent across sensors given the true object location $X$ and the sensing modes $A^{(m)}_n, n = 1, \ldots, N , m = 1, \ldots, M$.  This makes the error channel from the true indicators $Z^{(m)}_n$ to the measurements $Y^{(m)}_n$ memoryless. 

Based on the conditional independence assumptions, we define the conditional density of the joint measurements given the indicator variables $i_1, \ldots, i_M$ associated with the sensing modes and the state $X$, as 
\begin{align}
\label{eq:joint_error}
q_{i_{1:M}}(y^{(1)},\ldots,y^{(M)}) = \prod_{m=1}^{M} f_{i_m}^{(m)} (y^{(m)})
\end{align}
where we use $i_{1:M}$ as a shorthand for $(i_1,\ldots,i_M)$.

As before, we will collect observations in $N$ stages.  At each stage, as the sensors make decisions and obtain observations, the observed noisy values are shared and the posterior probability density of the object location $X$ will be updated for all sensors.  Note that the posterior density of $X$ is only updated after all the measurements are obtained and shared. 

Let $\boldsymbol{A}_n = (A^{(1)}_n, \ldots, A^{(M)}_n)$ denote the batch sensing modes used at stage $n$, and $\boldsymbol{Y}_n = (Y^{(1)}_n, \ldots, Y^{(M)}_n)$ denote the batch noisy observations collected at stage $n$.  The information history after stage $n$ is denoted by $D_n = \{ \boldsymbol{A}_1, \boldsymbol{Y}_1, \ldots, \boldsymbol{A}_n, \boldsymbol{Y}_n\}$.  Let the information  state after stage $n$ be denoted as $p_n(x)$, which is the posterior density 
$p_n(x) = p(x|p_0,D_n)$.  This information state evolves after collecting observations $\boldsymbol{Y}_{n+1} = \boldsymbol{y} = (y^{(1)}, \ldots, y^{(M)})$ for all $M$ sensors with batch sensing modes $\boldsymbol{A}_{n+1} = (A^{(1)}, \ldots, A^{(M)})$ as follows:

\begin{align*}  
p_{n+1}(x) &= p_n(x) \cdot \frac{P(\boldsymbol{Y}_{n+1}=\boldsymbol{y}|\boldsymbol{A}_{n+1}, X=x)}
{  \int_{\mathcal{X}} p_n(\sigma) P(\boldsymbol{Y}_{n+1}=\boldsymbol{y}|\boldsymbol{A}_{n+1}, X=\sigma) d\sigma           }\\
&=  p_n(x) \cdot \frac{    \sum_{i_1 = 0}^1\cdots \sum_{i_M=0}^1   q_{i_{1:M}}(y^{(1)},\ldots,y^{(M)})  \mathbbm{1}_{\{x \in \cap_{m=1}^M (A_n^{(m)})^{i_m}\}}      }
{  \int_{\mathcal{X}} p_n(\sigma) \sum_{i_1 = 0}^1\cdots \sum_{i_M=0}^1   q_{i_{1:M}}(y^{(1)},\ldots,y^{(M)})  \mathbbm{1}_{\{\sigma \in \cap_{m=1}^M (A_n^{(m)})^{i_m}\}} d\sigma           }
\end{align*}
where we use the notation $(B)^0 \equiv B^c$ and $(B)^1 \equiv B$ if $B$ is a subset of $\mathcal{X}$.

Define the collection of statistics, called the \textit{joint operating point},  as $\boldsymbol{u} = \{ u_{i_{1:M}}, i_1, \ldots, i_M \in \{0,1\} \}$ as 
\begin{align}
u_{i_{1:M}} = \int_{\cap_{m=1}^M (A^{(m)})^{i_m}} p_n(\sigma) d\sigma \ \geq 0
\label{eq:stats2}
\end{align}
Then,  $P(\boldsymbol{Y}_{n+1}=\boldsymbol{y}|\boldsymbol{A}_{n+1}, p_n) \equiv \eta(\boldsymbol{y}) $ can be evaluated as
\begin{align*}
\eta(\boldsymbol{y}) = \sum_{i_1=0}^1 \cdots \sum_{i_{M}=0}^1  q_{i_{1:M}}(y^{(1)},\ldots,y^{(M)})  u_{i_{1:M}},
\quad \mbox{ where }  \sum_{i_1=0}^1 \cdots \sum_{i_{M}=0}^1 u_{i_{1:M}} = 1
\end{align*}

The quality of our knowledge about the object position $X$ is evaluated by the {posterior differential entropy} $H(p_n)$ of the posterior density $p_n(x)$, defined as
\begin{align*}
H(p_n) = -\int_\mathcal{X} p_n(x) \log_2 p_n(x) dx.
\end{align*}
where $\mathcal{X}$ is the domain of $X$.

Let $S_n$ denote the space of probability densities $p_n(x)$ over $\mathcal{X}$.  Let $\Gamma(\mathcal{X})$ denote the set of all Lebesgue-measurable subsets of $\mathcal{X}$.  
We define an adaptive joint sensing policy $\pi = (\pi_1, \pi_2, \cdots, \pi_N)$ to be a sequence of functions where $\pi_n:S_{n-1} \rightarrow \Gamma(\mathcal{X})^M$ will map the posterior density $p_{n-1}(x)$ into  admissible batch sensing modes $\boldsymbol{A}_n = (A^{(1)}_n, \ldots, A^{(M)}_n)$. Let $\Pi$ denote the space of all adaptive joint sensing policies.  

Our goal is to minimize $H(p_N)$ --- the posterior differential entropy after  $N$ stages of joint sensing: 
\begin{align*}
\underset{\pi \in \Pi}{\inf}\;E[H(p_N) | p_0]
\end{align*}

To view this as a special case of our previous results, we simply need to recognize that a  set of joint sensing modes $\boldsymbol{A} = (A^{(1)}, \ldots, A^{(M)})$ induces a partition of the region $\mathcal{X}$ into $2^M$ subsets.  For each $k \in \{0, \ldots, 2^M - 1\}$, let $i_1, \ldots, i_M$ denote the dyadic expansion of $k$.  Then, the subset $k$ in the partition can be identified as
$$ A^k = \mathbbm{1}_{\{x \in \cap_{m=1}^M (A^{(m)})^{i_m}\}}$$
  Similarly, given any partition $A$ of the region $\mathcal{X}$ into $2^M$ subsets, we can use the dyadic expansion of $k$ to define joint sensing modes 
$$ A^{(m)} = \cup_{\{k : i_m = 1\}} A^k$$   By identifying the joint set of observations $(Y^{(1)}, \ldots, Y^{(M)})$ as a discrete valued observation $Y$ in a discrete-valued set $\mathcal{Y}^M$ with conditional probability distribution as \eqref{eq:joint_error}, we can map this problem into a special case of the multi-region single-sensor problem.  
Thus, the optimal solution for the Boolean multi-sensor problem is a special case of the results of Proposition \ref{prop:greedyIsOptThm_multianswer}.  This result has also been obtained directly for  Boolean multi-sensor search in  Theorem 1 of \cite{hero2014collaborative}.  We highlight this solution below.

One of the  quantities of interest is the expected entropy reduction of  batch sensing modes $\boldsymbol{A}_{n+1}$, corresponding to Proposition \ref{prop:OneStepEntropyUpdate_multianswer}.  This takes a special form for the multiple Boolean sensor case: 
\begin{align}
\label{eq:joint_phi}
\varphi(\boldsymbol{u}) = \mathcal{H}( \sum_{i_1=0}^1 \cdots \sum_{i_{M}=0}^1 q_{i_{1:M}} u_{i_{1:M}}) -  \sum_{i_1=0}^1 \cdots \sum_{i_{M}=0}^1 u_{i_{1:M}} \mathcal{H}(q_{i_{1:M}})
\end{align}
where $\boldsymbol{u}$ is defined in \eqref{eq:stats2} and $q_{i_{1:M}}(y^{(1)},\ldots,y^{(M)})$ is defined in \eqref{eq:joint_error}.

\begin{proposition}
\label{prop:OneStepEntropyUpdate_twosensor}
Given a batch of sensing modes for stage $n+1$ as  $\boldsymbol{A}_{n+1} = (A^{(1)}, \ldots, A^{(M)})$.  
\begin{align*}
H(p_n) - E[H(p_{n+1})|\boldsymbol{A}_{n+1} = (A^{(1)}, \ldots, A^{(M)}),p_n] = \varphi(\boldsymbol{u})
\end{align*}  
\end{proposition}

The strict concavity of the Shannon entropy $\mathcal{H}(f)$ can again be used to show $\varphi(\boldsymbol{u})$ is strictly concave and has a unique maximum at $\boldsymbol{u}^*$.  
Furthermore, we can always find a joint sensing strategy that achieves this optimal value using the correspondence identified above and the construction in the proof of Proposition \ref{prop:greedyIsOptThm_multianswer}.  

\begin{proposition}
\label{prop:multi_realizable}
At each stage $n$, we can always find a batch of sensing modes $\boldsymbol{A}_n = (A^{(1)}, \ldots, A^{(M)})$ such that  
\begin{align*}
\int_{\cap_{m=1}^M (A^{(m)})^{i_m}}  p_{n-1}(x) dx = u_{i_{1:M}}^*
\end{align*}  
\end{proposition}

These results can be used to establish the following result:

\begin{proposition}[Theorem 1 in \cite{hero2014collaborative}]
\label{prop:greedyIsOptThm_twosensor}
Let  $\boldsymbol{u}^* = \arg\max_{\boldsymbol{u}} \varphi(\boldsymbol{u})$ for $\varphi(\boldsymbol{u})$ as defined in \eqref{eq:joint_phi}. For each stage $n$, one can select batch sensing modes $\boldsymbol{A}_n = (A^{(1)}, \ldots, A^{(M)})$ that satisfy $\boldsymbol{u}(\boldsymbol{A}_n, p_{n-1}) = \boldsymbol{u}^*$. Then, this adaptive set of policies is optimal for the multi-sensor joint sensing problem.  Furthermore, the optimal value function is given by
\begin{align}
\label{eq:value_fun_multisensor}
V(p_n,n)=H(p_n) - (N-n)\varphi^*
\end{align}
where  $\varphi^* = \varphi(\boldsymbol{u}^*)$.
\end{proposition}

Note that obtaining $\boldsymbol{u}^*$ as indicated above for general discrete error models requires solution of a large concave maximization problem (with $2^M$ variables), which can be time-consuming.  We show next that we can obtain this optimal solution through the solution of $M$ scalar convex optimization problems, which is a much simpler problem.    

\begin{figure}[htb]
  \centering
  \includegraphics[scale=0.27]{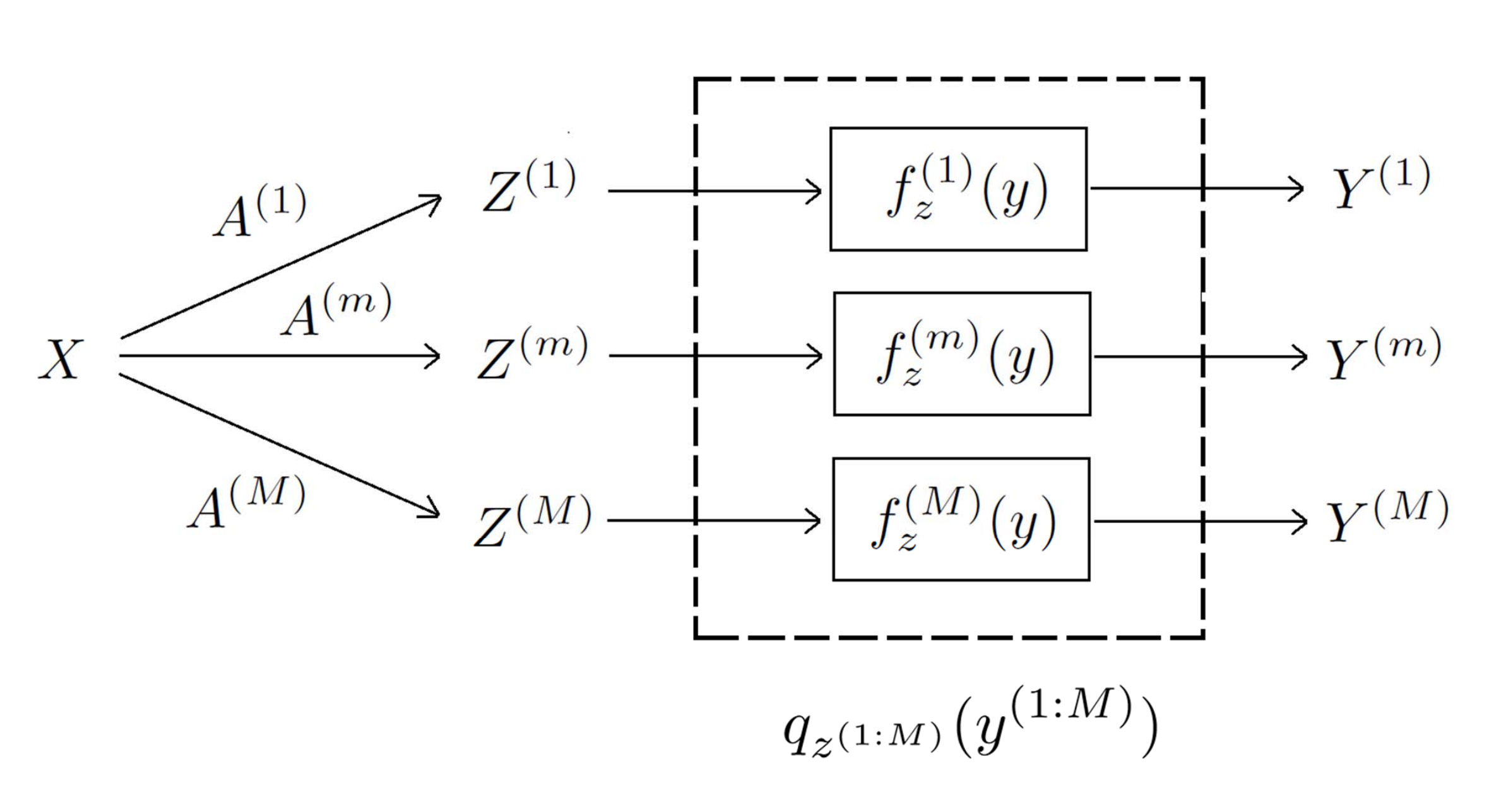}
  \caption{The ``mixed' channel.}
  \label{fig:mixedChannel}
\end{figure}

One way to view the results of Proposition \ref{prop:greedyIsOptThm_twosensor} is to connect the maximal entropy reduction at each stage to the concept of channel capacity.  Each sensor $m$ can be viewed as a discrete memoryless stationary channel (DMSC) whose input is $Z^{(m)} \in \{0,1\}$, output is $Y^{(m)} \in \mathcal{Y}$, and transition probabilities are specified by $f_1^{(m)}$ and $f_0^{(m)}$.  Furthermore, we can regard $g_{i_{1:M}}(y^{(1)},\ldots,y^{(M)})$ defined in \eqref{eq:joint_error} as the transition probabilities of a ``mixed'' vector (product) channel for all the $M$ sensors used in joint sensing, shown in Fig. \ref{fig:mixedChannel}.  The inputs of this mixed channel are vector $(Z^{(1)},\ldots,Z^{(M)}) \in \{0,1\}^M$, and the outputs are $(Y^{(1)},\ldots,Y^{(M)}) \in \mathcal{Y}^M$. The capacity  $C_{mix}$ of this channel is equal to $\varphi^*$, the solution of our optimization problem in Proposition \ref{prop:greedyIsOptThm_twosensor}.  

Consider now the problem when there is only one sensor present, as in \cite{20qs}.  Define $\varphi^{(m)}(u^{(m)})$ to be the single sensor expected differential entropy reduction that sensor $m$ can achieve on its own by selecting its sensing mode $A^{(m)}$, as 
\begin{align*}
\varphi^{(m)}(u^{(m)}) = \mathcal{H}(u^{(m)} f_1^{(m)} + (1-u^{(m)}) f_0^{(m)})
 - u^{(m)} \mathcal{H}(f_1^{(m)}) - (1-u^{(m)}) \mathcal{H}(f_0^{(m)}) \  
\end{align*}
and define the maximum expected entropy reduction to be 
\begin{align*}
\varphi^{(m)*} = \max_{u^{(m)}} \varphi^{(m)}(u^{(m)}) 
\end{align*}

The following proposition is proved in the Appendix:

\begin{proposition}
\label{thm:equiv}
Consider general discrete-output sensor error models. Denote the optimal operating points of each individual sensor $m$ as $u^{(m)*} = \arg\max_{u^{(m)}} \varphi^{(m)}(u^{(m)})$, $m=1,\ldots,M$.
  Then the optimal operating point for joint sensing, i.e., $\boldsymbol{u}^* = (u_{i_{1:M}}^*) = \operatorname*{arg\,max}_{\boldsymbol{u}} \varphi(\boldsymbol{u})$, is given by
\begin{align}   \label{eq:uopt_from_ufug}
u_{i_{1:M}}^* = \prod_{m=1}^M (u^{(m)*})^{i_m} (1-u^{(m)*})^{1-i_m}
\end{align}
In addition, 
\begin{align*}
\varphi^*   =   \sum_{m=1}^M  \varphi^{(m)*}  
\end{align*}
\end{proposition}

Thus, the optimal joint operating point for the multiple Boolean sensor case with discrete measurements can obtained from the optimal single-sensor operating points.  Furthermore, we can now use the construction of Section \ref{subsec:optimal_policy_multianswer} to obtain partitions of the region $\mathcal{X}$ that achieve the probabilities required by the joint operating point, and combine them to obtain the joint optimal sensing modes for each sensor $m \in \{1, \ldots, M\}$.  

The results of Proposition \ref{thm:equiv} can also be used to identify an equivalence between joint sensing and sequential sensing for general Boolean sensors.  The authors in  \cite{hero2014collaborative} propose a sequential sensing scheme where each stage is divided into $M$ sub-stages.  At each substage $m$, the $m$-th sensor selects a sensing mode based on information state $p_{n,m-1}$, where $p_{n,0} \equiv p_n(x)$, and collects its noisy measurement.  This measurement is processed to obtain an updated probability density for the object location $p_{n,m}(x)$.  This information is made available to the next sensor $m+1$, which in turn selects its query based on $p_{n,m}(x)$.  The stage completes when the $M$-th sensor collects its measurement, and uses it to update $p_{n,M-1}(x)$ to produce $p_{n+1}(x)$.  

This sequential sensing scheme is similar to the binary single-sensor search problem studied in \cite{20qs}, with the minor extension that the sensor error models used for each substage are not time invariant.  From \cite{20qs} we know that the optimal policies are the ones that select a sensing mode to maximally reduce the expected posterior entropy in each single substage.  Thus, the optimal expected differential entropy reduction for the sequential policy at the end of one cycle is precisely 
 $\sum_{m=1}^M  \varphi^{(m)}(u^{(m)}) $, 
which is the same value  $\varphi^*$ that would be achieved by the joint sensing scheme in  Proposition \ref{thm:equiv}.  This establishes the following lemma for general discrete observation Boolean sensors:

\begin{lemma}
\label{lemma:comparison}
Consider general discrete-output sensor error models for both the sequential and joint sensing models described above.  Then, the optimal performance achievable at the end of a stage is the same for both sequential and joint sensing.  
\end{lemma}

The above lemma extends the results of  \cite{hero2014collaborative} to Boolean sensors with general discrete error models and non-binary measurements.  Note also that we can shuffle arbitrarily the orders in which sensors take measurements in a substage and achieve the same result.  

To illustrate these results, consider a problem with two Boolean sensors $f$ and $g$ whose error models are specified in Table \ref{table:BAC_sensor_model}.  For each individual sensor, the optimal operating points correspond to the points that achieve maximal capacity in an asymmetric binary channel (e.g. \cite{asymcap}), and are given by:

\begin{small}
\begin{align*}
u^{(f)*} &= \frac{\rho(1+k_1)-1}{(\alpha + \rho -1)(1+k_1)}\\
u^{(g)*} &= \frac{\Delta(1+k_1)-1}{(\beta + \Delta -1)(1+k_1)}
\end{align*}
\end{small}
where
\begin{small}
\begin{align*}
k_1 =  \Big( \frac{ \alpha^\alpha (1-\alpha)^{(1-\alpha)} }{ \rho^\rho (1-\rho)^{(1-\rho)} } \Big)^{\frac{1}{\alpha+\rho-1}},   \quad
k_2 =  \Big( \frac{\beta^\beta (1-\beta)^{(1-\beta)}  }{\Delta^\Delta (1-\Delta)^{(1-\Delta)}  } \Big)^{\frac{1}{\beta+\Delta-1}}
\end{align*}
\end{small}

\begin{table}[tb]
    \begin{minipage}{.5\linewidth}
      \centering
        \begin{tabular}{c || c | c}
          \hline
                     & $y=1$      & $y=0$      \\ \hline
            $f_1(y)$ & $\alpha$   & $1-\alpha$ \\ \hline
            $f_0(y)$ & $1-\rho$  & $\rho$  \\         
          \hline
        \end{tabular}
        \caption*{(a) Sensor $f$. ($\alpha  \neq \rho$)}
    \end{minipage}%
    \begin{minipage}{.5\linewidth}
      \centering
        \begin{tabular}{c || c | c}
          \hline
                      & $y'=1$     & $y'=0$     \\ \hline
            $g_1(y')$ & $\beta$    & $1-\beta$  \\ \hline
            $g_0(y')$ & $1-\Delta$ & $\Delta$   \\           
          \hline
        \end{tabular}
        \caption*{(b) Sensor $g$. ($\beta   \neq \Delta$)}
    \end{minipage}
    \caption{Two Boolean Sensors with Non-Symmetric Binary-Output Error Models}
    \label{table:BAC_sensor_model}
\end{table}

We can combine these solutions as in Proposition \ref{thm:equiv} to obtain the following joint operating points: 
\begin{small}
\begin{align*}
u_{i_1i_2}^* 
= \frac{   (\rho(1+k_1)-1)^{i_1}   (\alpha(1+k_1)-k_1)^{1-i_1}   }{(\alpha+\rho-1)(1+k_1)}
\cdot
\frac{   (\Delta(1+k_2)-1)^{i_2}    (\beta(1+k_2)-k_2)^{1-i_2}   }{(\beta+\Delta-1)(1+k_2)},
\; i_1, i_2 \in \{0,1\}
\end{align*}
\end{small}

There is a further simplification where the optimal operating points for the joint sensing problem can be computed analytically. The results in  \cite{hero2014collaborative} show that, when each of the $M$ sensors has a binary symmetric error model, the optimal operating points are $u_{i_{1:M}}^* = \frac{1}{2^M}$, and the optimal individual sensor operating points are $u^{(m)*} = 1/2$.  We extend this further to non-binary error models where $|\mathcal{Y}| \geq 2$.  Specifically, we consider the case of Boolean sensors with error models that correspond to quasi-symmetric memoryless channels \cite{Chen00} (also called symmetric discrete memoryless channels in \cite{Gallager}.  For Boolean sensors, quasi-symmetric error models satisfy the following  condition:

\begin{definition}[Symmetry condition]
A Boolean sensor has a quasi-symmetric error model if there exists a permutation $\chi():\mathcal{Y}\rightarrow \mathcal{Y}$ such that, for all $y \in \mathcal{Y}$,  $f_1 (y) = f_0 (\chi(y))$ and $f_0(y)= f_1(\chi(y))$.  
\end{definition}

The implications of having quasi-symmetric  error models are:
\begin{align*}
\mathcal{H}(f_1) &= \mathcal{H}(f_0) \\
  \sum_{y \in \mathcal{Y}} (f_1(y)) \log \frac{f_1(y) + f_0(y)}{2}
&= \sum_{y \in \mathcal{Y}} (f_0(y)) \log \frac{f_1(y) + f_0(y)}{2} 
\end{align*}

These lead to the well-known result \cite{Chen00,Gallager} that the optimal capacity for quasi-symmetric discrete memoryless channels is achieved by a uniform input distribution.  For Boolean sensors, this means that the optimal sensing mode at stage $n$ with information state $p_{n-1}$ is to pick a subset $A_n \in \mathcal{X}$ such that $\int_{x \in A_n} p_{n-1}(x) dx = 1/2.$  
This establishes the following result as a direct application of Proposition \ref{thm:equiv}:
\begin{corollary}
\label{lemma:sym_equal_partition}
If we have $M$ Boolean sensors with quasi-symmetric (albeit different) error models, then the maximum value of $\varphi(\boldsymbol{u})$ occurs at $u_{i_{1:M}}^* = 2^{-M}$.
\end{corollary}

When the error model is a discrete-continuous channel,  $\mathcal{Y}$ is a subset of the real line and $f_0(y), f_1(y)$ are probability densities absolutely continuous with respect to Lebesgue measure.  In order for a single Boolean sensor to achieve its optimal capacity with uniform input distribution $u = 0.5, 1-u = 0.5$, we must satisfy the following conditions:
\begin{align*}
\mathcal{H}(f_1) &= \mathcal{H}(f_0) \\
  \int_{y \in \mathcal{Y}} (f_1(y)) \log \frac{f_1(y) + f_0(y)}{2} dy
&= \int_{y \in \mathcal{Y}} (f_0(y)) \log \frac{f_1(y) + f_0(y)}{2} dy 
\end{align*}
When these conditions are satisfied, the derivative of $\varphi$ with respect to $u$ is:
$$\frac{\partial \varphi}{\partial u} = - \int_{y \in \mathcal{Y}} (f_1(y) - f_0(y)) \log(u f_1(y) + (1-u) f_0(y)) dy - H(f_1) + H(f_0)$$
This derivative vanishes at $u = 0.5$ when the above conditions are satisfied.  Coupled with strict concavity in $u$ of the $\varphi$ function, this leads to the optimal operating point for  single sensor with discrete-continuous error model to be at $u = 0.5$.  

A sufficient condition on the densities $f_1(y)$, $f_0(y)$ to satisfy the above conditions is that  there exists some constant $\alpha$ such that $f_0(y) = f_1(\alpha - y)$ for all $y$.  In this case, the differential entropies $H(f_1)$ and $H(f_0)$ are equal, and the equality 
$$  \int_{y \in \mathcal{Y}} (f_1(y)) \log \frac{f_1(y) + f_0(y)}{2} dy
= \int_{y \in \mathcal{Y}} (f_0(y)) \log \frac{f_1(y) + f_0(y)}{2} dy $$
is easily verified.  These conditions are satisfied when the error models correspond to additive white Gaussian channels, so that 
$$ Y_n = h(Z_n) + w_n$$
where $h(Z_n)$ is a function of the Bernoulli variable $Z_n$ and $w_n$ is zero-mean with Gaussian distribution.  

\section{Boolean Multi-Sensor with Precision Modes Selection}
\label{subsec:precision_mode}
In \cite{microscopy}, Sznitman et al. generalized \cite{20qs} to the setting where at each sensing stage, the Boolean sensor is allowed to choose a \textit{precision mode} from a finite number of precision modes, in addition to its observed subset.   A precision mode for a sensor selects a particular error model, and there is a cost associated with selection of different precision modes.  Different precision modes will trigger different sensor error models, but there are also costs associated with the precision modes.   Precision modes with better error models will cost more to use, but may result in greater reduction in differential entropy.   In this subsection, we will extend this to the Boolean multi-sensor joint search problem, allowing each sensor to choose its precision mode at each stage.

Assume that for the $m$-th sensor, the set of its possible precision modes is $\mathcal{L}^{(m)} = \{1,\ldots,L^{(m)}\}$.  A joint search decision at stage $n$ consists of selecting both sensing modes and precision modes for all $M$ sensors $(A_n^{(1)}, l_n^{(1)}, \ldots, A_n^{(M)}, l_n^{(M)})$, where $A_n^{(m)} \subset \mathcal{X}$ and $l_n^{(m)} \in \mathcal{L}^{(m)}$ for $\forall m$.

The corresponding error models of the $m$-th sensor under precision mode $l_n^{(m)} = l$ is characterized by $f_1^{(m,l)}$ and $f_0^{(m,l)}$, defined as:
\begin{small}
\begin{align*}
P(Y_n^{(m)} = y | A_n^{(m)} = A, l_n^{(m)} = l, X)  =
\begin{cases}
f_1^{(m,l)} (y)  \quad  &  \mbox{if }  X \in A   \\
f_0^{(m,l)} (y)  \quad  &  \mbox{otherwise}
\end{cases}
\end{align*}
\end{small}

When sensor $m$ selects a precision mode $l_n^{(m)} = l$, it incurs a cost $W^{(m)}(l)$.  Note that this cost depends on the sensor index $m$ and the precision mode $l$, but does not depend on the sensing mode $A$ or the time instance $n$ (although this time invariance restriction can be easily relaxed).  

Define the information state $p_n(x) = p(x | (\boldsymbol{A}_1, \boldsymbol{l}_1), \boldsymbol{Y}_1, \ldots, (\boldsymbol{A}_n, \boldsymbol{l}_n), \boldsymbol{Y}_n)$.  The state evolves according to Bayes rule as before, where the choices of precision modes are used to select the appropriate likelihoods for interpreting  the observed measurements $\boldsymbol{Y}$.  A joint sensing policy $\pi = (\pi_1, \ldots, \pi_N)$ with precision modes selection is a sequence of functions that map the information state $p_{n-1}(x)$ to admissible batch sensing and precision modes $(\boldsymbol{A}_n, \boldsymbol{l}_n) = (A_n^{(1)}, l_n^{(1)}, \ldots, A_n^{(M)}, l_n^{(M)})$ at each stage $n$.   Denote the policy space by $\Pi$.   

Our joint sensing objective is to minimize the sum of the final-stage expected posterior differential entropy and the total cost of all sensors discounted by a factor $\gamma$.  The resulting :
\begin{align*}
\inf_{\pi \in \Pi}  E[H(p_N) + \gamma \sum_{t=1}^N \sum_{m=1}^M W^{(m)}(l_t^{(m)}) | p_0]
\end{align*}

This stochastic control problem fits the countable disturbance model of \cite{Ber_Shreve}.  Define the optimal value function $V(p_n, n)$ at stage $n$ to be:
\begin{align*}
V(p_n, n) = \inf_{(\pi_{n+1}, \ldots, \pi_{N})} E [H(p_N) + \gamma   \sum_{t=n+1}^N \sum_{m=1}^M W^{(m)}(l_t^{(m)})  | p_n]
\end{align*}

The Bellman equation for this problem is:
\begin{align*}
V(p_n, n) = \inf_{\boldsymbol{A}_{n+1}, \boldsymbol{l}_{n+1}} E_{\mathbf{Y}_{n+1}}[V(p_{n+1}, n+1) + \gamma \sum_{m=1}^M W^{(m)}(l_{n+1}^{(m)})  | (\boldsymbol{A}_{n+1}, \boldsymbol{l}_{n+1}), p_n]
\end{align*}
If a policy $\pi \in \Pi$ achieves equality in the Bellman equation, it is an optimal policy.  

Following our previous approach, consider a set of sensing decisions $(\boldsymbol{A}_{n+1}, \boldsymbol{l}_{n+1})$ at stage $n+1$.   Define the joint operating point $\boldsymbol{u} = \{ u_{i_{1:M}}, i_1, \ldots, i_M \in \{0,1\} \}$ as 
\begin{equation}
\label{eq:pstats}
u_{i_{1:M}} = \int_{\cap_{m=1}^M (A^{(m)})^{i_m}} p_n(\sigma) d \sigma \ge 0
\end{equation}

For a given set of sensing decisions at stage $n+1$, consider the one-stage gain to be the expected reduction in differential entropy minus the cost of the precision modes by the sensors, as 
$$\hat{G}(p_n,\boldsymbol{A}_{n+1}, \boldsymbol{l}_{n+1}) \equiv   \mathcal{H}(p_n)  - E_{\mathbf{Y}_{n+1}}[\mathcal{H}(p_{n+1}) + \gamma \sum_{m=1}^M W^{(m)}(l_{n+1}^{(m)})  | (\boldsymbol{A}_{n+1}, \boldsymbol{l}_{n+1}), p_n] $$

We have the following characterization:
\begin{lemma}
\label{lemma:oneStage_precisionMode}
Define the function 
\begin{small}
\begin{align*}
G(\boldsymbol{u}, \boldsymbol{l}) = \mathcal{H}(\sum_{i_1=0}^1 \cdots \sum_{i_M = 0}^1 q_{i_{1:M}}^{\boldsymbol{l}} u_{i_{1:M}}) - \sum_{i_1=0}^1 \cdots \sum_{i_M = 0}^1 u_{i_{1:M}} \mathcal{H}(q_{i_{1:M}}^{\boldsymbol{l}}) - \gamma  \sum_{m=1}^M W^{(m)}(l^{(m)})
\end{align*}
\end{small}
where $\boldsymbol{u}(\cdot)$ is  defined from $p_n$ and $\boldsymbol{A}$ as in \eqref{eq:pstats} and
$$q_{i_{1:M}}^{\boldsymbol{l}}(y^{(1)},\ldots,y^{(m)}) = \prod_{m=1}^M f_{i_m}^{(m, l^{(m)})}(y^{(m)}), y^{(m)} \in \mathcal{Y}, i_m \in \{0,1\}, \forall m$$

Then, for all sensing decisions, the one-stage gain can be expressed  as
\begin{align*}
\hat{G}(p_n,\boldsymbol{A}_{n+1}, \boldsymbol{l}_{n+1}) &\equiv G( \boldsymbol{u}_{n+1},  \boldsymbol{l}_{n+1})
\end{align*} 
\end{lemma}

Thus, the dependence of the one stage gain on the information state and the selected sensing modes is summarized by the statistics $\boldsymbol{u}$.  This structure is exploited to derive the main result:

\begin{proposition}
\label{prop:optimal_precision_mode}
Consider the problem of finding the optimal sensing modes and precision modes for the Boolean multisensor problem.  Define  $(\boldsymbol{u}^*, \boldsymbol{l}^*) \in \operatorname*{arg\,max}_{(\boldsymbol{u},\boldsymbol{l})} G(\boldsymbol{u},\boldsymbol{l})$.  Then, at stage $n$, any policy that chooses batch sensing and precision modes $(\boldsymbol{A}_n, \boldsymbol{l}_n)$ such that $\boldsymbol{l}_n = \boldsymbol{l}^* = (l^{(1)*}, \ldots, l^{(M)*})$ and $\boldsymbol{u}(\boldsymbol{A}_n, p_{n-1}) = \boldsymbol{u}^*$ is optimal.   

Furthermore, the optimal value function has the following closed-form expression:
\begin{align*}
V(p_n,n) = H(p_n)  - (N-n) G^*
\end{align*}
where $G^* = G(\boldsymbol{u}^*, \boldsymbol{l}^*)$.
\end{proposition}

Note that, for each $\boldsymbol{l}$, the function $G(\boldsymbol{u},\boldsymbol{l})$ is strictly concave in $\boldsymbol{u}$.  Thus, we can find the maximum value and maximizing argument for each $\boldsymbol{l}$, and then select the maximum value among the possible choices of $\boldsymbol{l}$ to obtain $(\boldsymbol{u}^*, \boldsymbol{l}^*)$.  The maximizing argument may not be unique, because there can be multiple precision modes with the same maximum value.  As long as the sensor error models are stage-invariant, there is an optimal strategy where each sensor will choose the same sensor-dependent precision mode at every sensing stage.   

We now show that optimal strategies for the solution of the optimal Boolean multi-sensor problem with precision mode selection can be computed from the solution of single sensor problems:  Let
\begin{align*}
G^{(m)}(u, l) = \mathcal{H}( u f^{(m,l)}_1(y) + (1-u) f^{(m,l)}_0(y) ) - u \mathcal{H}(f^{(m,l)}_1(y)) - (1-u)\mathcal{H}(f^{(m,l)}_0(y))  - \gamma  W^{(m)}(l)
\end{align*}
and let $(u^{(m)*}, l^{(m)*}) = \operatorname*{arg\,max}_{u,l} G^{(m)}(u, l)$ denote an optimal operating point and choice of precision mode for sensor $m$.  

\begin{proposition}
\label{thm:equiv_prec}
Let $ (u^{(m)*}, l^{(m)*}) = \operatorname*{arg\,max}_{u,l} G^{(m)}(u, l)$ denote an optimal operating point and choice of precision mode for each sensor $m = 1, \ldots, M$.  Then, an optimal joint operating point and precision modes for joint sensing $ (\boldsymbol{u}^*, \boldsymbol{l}^*)$ can be obtained as:
\begin{align}   \label{eq:full_from_sing}
u_{i_{1:M}}^* = \prod_{m=1}^M (u^{(m)*})^{i_m} (1-u^{(m)*})^{1-i_m}; \quad \boldsymbol{l}^* = (l^{(1)*}, \ldots, l^{(M)*})
\end{align}
In addition, 
\begin{align*}
G^*   =   \sum_{m=1}^M  G^{(m)*}  
\end{align*}
where $ G^{(m)*} = G^{(m)}(u^{(m)*}, l^{(m)*})$.
\end{proposition}

\section{Simulation Results of Finite Horizon Boolean Multi-Sensor Joint Search}
\label{sec:simulation_multisensor}

\subsection{Two Boolean Sensors with Binary Symmetric or Gaussian Error Models}
\label{subsection:simulation_twosensor}

In this subsection, we illustrate the previous results by simulating a two sensor scenario searching for a object.  We assume the object is located in the unit interval $[0,1]$, i.e., $\mathcal{X}=[0,1]$. We denote the two sensors as $f$ and $g$, who will cooperate and share their sensed measurements to gain the knowledge of the object position $X$.  The sensing modes of $f$ and $g$ at stage $n$ are denoted as $A_n^f$ and $A_n^g$ respectively.

In terms of sensing errors, we consider two types of models: In the first model, $\mathcal{Y}$ is binary valued, with symmetric errors for each sensor, but with different probabilities of error.  The measurement probability distribution functions $f_0(y), f_1(y), g_0(y)$ and $g_1(y)$ are summarized in Table \ref{table:sensorSpecs}(a).  

For the second model, we assume that $y$ is continuous valued.  The measurement $y$ corresponding to a sensing mode $A^f_n$ for sensor $f$ is given as
$$ y^f_n = \mathbbm{1}_{\{X \in A^f_n\}} + w^f_n$$
where $w^f_n$ is a Gaussian random variable, mean 0, variance 1.  A similar model is used for sensor $g$, with the Gaussian random variables independent across sensors and stages.  As required in our model, the additive noises are independent across stages, and between sensors.  The resulting probability densities are summarized in Table \ref{table:sensorSpecs}(b). 

\begin{table}[!htb]
    \caption{The sensor specifications for two types of Boolean sensors using optimal joint sensing policy. (a) Both sensors have symmetric error models, so $u_{11}^* =\cdots=u_{00}^*=\frac{1}{4}$. (b) Both sensors have Gaussian error models that satisfy symmetry, so $u_{11}^* =\cdots=u_{00}^*=\frac{1}{4}$.}
    \begin{minipage}{.5\linewidth}
      \centering
        \begin{tabular}{c || c | c}
          \hline
                     & $y=0$ & $y=1$ \\ \hline
            $f_0(y)$ & 0.8   & 0.2   \\ \hline
            $f_1(y)$ & 0.2   & 0.8   \\ \hline
            $g_0(y)$ & 0.7   & 0.3   \\ \hline
            $g_1(y)$ & 0.3   & 0.7   \\ \hline 
        \end{tabular}
        \caption*{(a) Binary symmetric error models}
    \end{minipage}%
    \begin{minipage}{.5\linewidth}
      \centering
        \begin{tabular}{c || c }
          \hline
                     & $y \in \mathbb{R}$       \\ \hline
            $f_0(y)$ & $y\sim\mathcal{N}(0,1)$  \\ \hline
            $f_1(y)$ & $y\sim\mathcal{N}(1,1)$  \\ \hline
            $g_0(y)$ & $y\sim\mathcal{N}(0,1)$  \\ \hline
            $g_1(y)$ & $y\sim\mathcal{N}(1,1)$  \\ \hline       
        \end{tabular}
        \caption*{(b) Gaussian error models}
    \end{minipage}
    \label{table:sensorSpecs}
\end{table}

Given the sensor models in Table \ref{table:sensorSpecs}, the first step is to find the operating points that minimize the $\varphi(\boldsymbol{u})$ as defined in proposition \ref{prop:OneStepEntropyUpdate_twosensor}.  In general, this would require a convex optimization problem, but the sensor models satisfy the  symmetry conditions discussed in Section \ref{sec:multisensor}, so the optimal values are $\boldsymbol{u}^*=(\frac{1}{4}, \frac{1}{4}, \frac{1}{4}, \frac{1}{4})$ (by Corollary \ref{lemma:sym_equal_partition}).

The optimal strategies at each stage $n$, based on the information state $p_{n-1}(x)$, is to find regions $A^f_n, A^g_n \subset [0,1]$ so that 
\begin{align*}
&\int_{x \in A^f_n \cap A^g_n} p_{n-1}(x) dx = 1/4\\
&\int_{x \in A^f_n \cap (A^g_n)^c} p_{n-1}(x) dx = 1/4\\
&\int_{x \in (A^f_n)^c \cap A^g_n} p_{n-1}(x) dx = 1/4\\
&\int_{x \in (A^f_n)^c \cap (A^g_n)^c} p_{n-1}(x) dx = 1/4
\end{align*}

\begin{figure}[htb]
  \centering
  \includegraphics[scale=0.5]{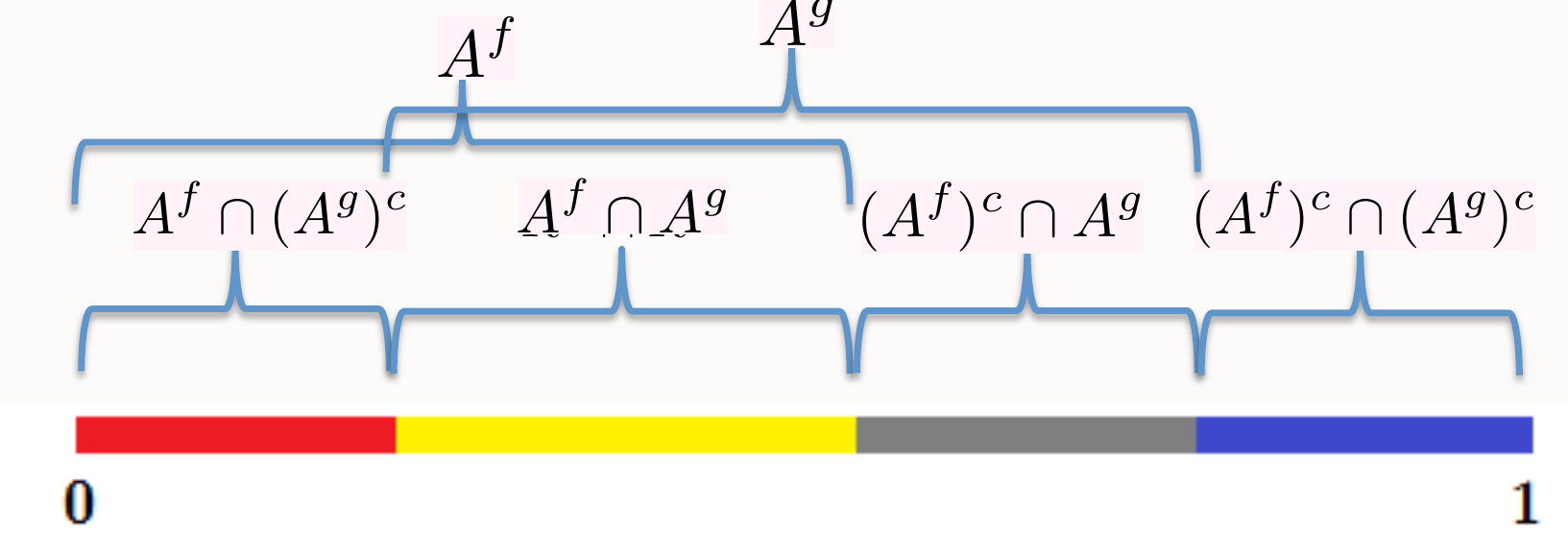}
  \caption{Partition of a line segment into four disjoint subsets at each stage. The domain $\mathcal{X}$ of the object position is the line segment $[0,1]$. At each stage, sensor $f$ will inquire subset $A^{f}$ and sensor $g$ will inquire subset $A^{g}$, thus partitioning $\mathcal{X}$ into four disjoint subsets.}
  \label{fig:linePartition}
\end{figure}

While there are many subsets that can satisfy the above equalities, we choose our subsets to be subintervals, to resemble physical properties of sensors that will likely observe connected regions.  Hence, we will partition $\mathcal{X}=[0,1]$ into four subintervals, each of which has probability 1/4 according to $p_{n-1}(x)$, corresponding to $A^f_n \cap (A^g_n)^c, A^f_n \cap A^g_n, (A^f_n)^c \cap A^g_n, (A^f_n)^c \cap (A^g_n)^c$.  This construction is illustrated in Fig. \ref{fig:linePartition}.  The subintervals are then used to identify the sensing modes $A^f_n, A^g_n$ for each sensor at stage $n$.  

\begin{figure*}[ht]
\centering
\begin{minipage}{.4\textwidth}
  \centering
  \includegraphics[width=\linewidth]{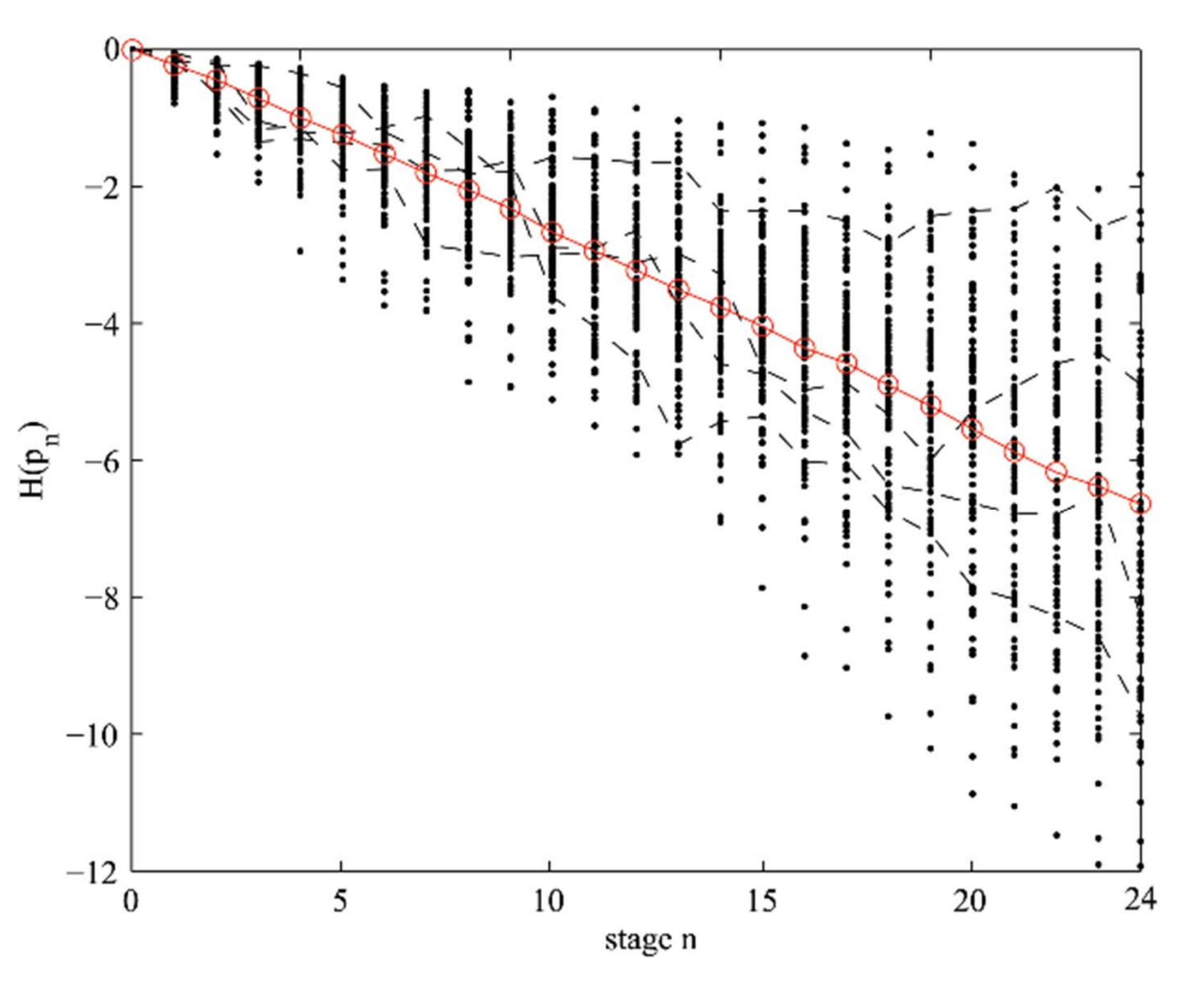}
  \caption*{(a) Binary symmetric error models}
  \label{fig:sub1}
\end{minipage}%
\begin{minipage}{.4\textwidth}
  \centering
  \includegraphics[width=\linewidth]{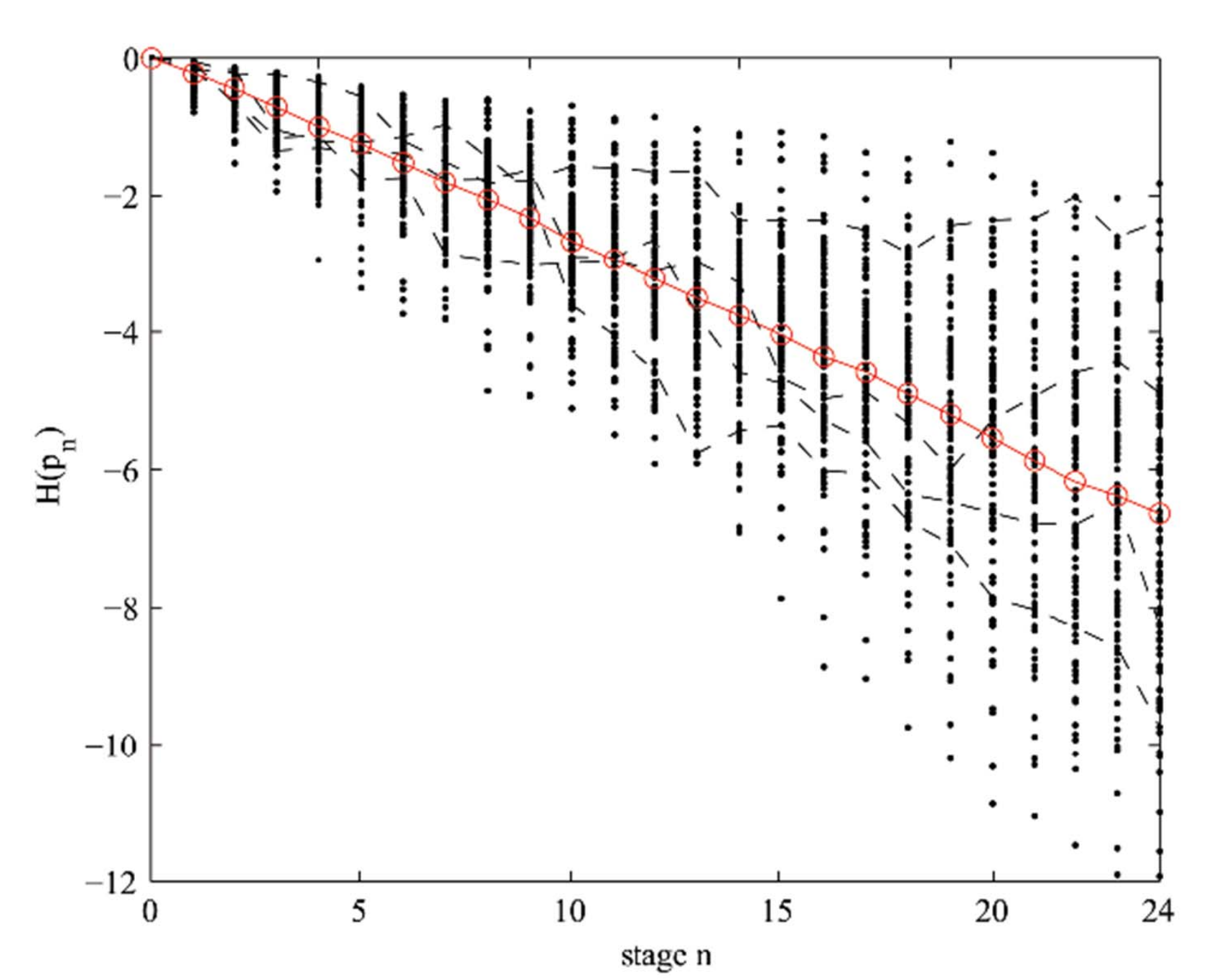}
  \caption*{(b) Gaussian error models}
  \label{fig:sub2}
\end{minipage}
\caption{The entropy reduction paths for two types of sensors. The red line in each figure is the average entropy reduction path over 100 realizations. We can see that the average entropy reduction paths for both types of sensors are a straight line, i.e., the average entropy decreases linearly. (a) Both sensors have binary symmetric error models, as described in Table \ref{table:sensorSpecs} (a), and we obtain that  $u_1^*=\cdots=u_4^*=\frac{1}{4}$. (b) Both sensors have Gaussian error models, as described in Table \ref{table:sensorSpecs} (b), and we obtain that $u_1^*=\cdots=u_4^*=\frac{1}{4}$.}
\label{fig:entropyReduction}
\end{figure*}

For each of the sensor models, we conducted 100 Monte Carlo experiments. In each experiment, we randomly generate a object position $X \in \mathcal{X}=[0,1]$ using a uniform distribution. We initialize our prior density for $X, p_0(x)$ as a uniform distribution; therefore, the initial differential entropy  $H(p_0) = -\int_0^1  \log_2(1) dx = 0$.  At each stage $n>0$, given the density $p_{n-1}(x)$, sensing modes $A^f_n, A^g_n$ are selected, and random measurements $(y^f_n,y^g_n)$ are generated according to the sensor error models.  These measurements are used to update the conditional density from $p_{n-1}(x)$ to $p_n(x)$ as indicated in Subsection \ref{subsec:multisensor_model}.  We continue this process until $n = 24$ sensing stages are completed.  

For each experiment, we plot the differential entropy $H(p_n)$ as a function of $n$.  Fig. \ref{fig:entropyReduction}(a) contains the results for the discrete measurement sensor model.  At each stage $n$, the plot displays the 100 sample values of $H(p_n)$.  The plot also shows four sample trajectories of experiments as dashed lines, to show the randomness in the actual trajectories.  Note that the sample trajectories of posterior differential entropy are not monotonically decreasing, as they depend on random measurement values.  The figure also shows in red the average of the 100 samples, which follows the linear descent predicted by the optimal value function in Proposition \ref{prop:greedyIsOptThm_twosensor}.

Fig. \ref{fig:entropyReduction}(b) shows similar results for the continuous measurement sensor model.  The vertical scales of the two plots are different.  The average slope of the posterior differential entropy decays slower in this graph, and the distribution of potential values has more support on higher values of differential entropy.  The graph still shows the expected linear decay from Proposition \ref{prop:greedyIsOptThm_twosensor}.

\subsection{Three Boolean Sensors with Ternary-Output Error Models}
\label{subsection:simulation_threesensor}

In this subsection, we simulate searching for a object in $\mathcal{X}=[0,1]$ using three sensors jointly, denoted as $f$, $g$ and $h$.  The sensing modes of $f$, $g$ and $h$ at stage $n$ are denoted as $A_n^f$, $A_n^g$ and $A_n^h$ respectively.

The sensor error model we consider here is shown in Table \ref{table:sensorSpecs_3sensor}.   Note that the error model of each sensor satisfies the symmetry conditions; thus, the optimal joint operating point is
\begin{small}
\begin{align*}
\boldsymbol{u}^* = (u_{000}^*, u_{001}^*, u_{011}^*, u_{010}^*, u_{110}^*, u_{111}^*, u_{101}^*, u_{100}^*) = (\frac{1}{8}, \frac{1}{8}, \frac{1}{8}, \frac{1}{8}, \frac{1}{8}, \frac{1}{8}, \frac{1}{8}, \frac{1}{8}).
\end{align*}
\end{small}

\begin{table}[!htb]
    \caption{The error models for three Boolean sensors}
    \centering
    \begin{tabular}{c || c | c | c}
        \hline
                     & $y=0$ & $y=1$ & $y=2$ \\ \hline
            $f_0(y)$ & 0.3   & 0.5   & 0.2   \\ \hline
            $f_1(y)$ & 0.2   & 0.5   & 0.3   \\ \hline
            $g_0(y)$ & 0.7   & 0.2   & 0.1   \\ \hline
            $g_1(y)$ & 0.2   & 0.7   & 0.1   \\ \hline       
            $h_0(y)$ & 0.3   & 0.1   & 0.6   \\ \hline
            $h_1(y)$ & 0.3   & 0.6   & 0.1   \\ \hline  
    \end{tabular}
    \label{table:sensorSpecs_3sensor}
\end{table}

The optimal joint sensing strategies at each stage $n$, based on the information state $p_{n-1}(x)$, is to find regions $A_n^f$, $A_n^g$, $A_n^h$ $\subset \mathcal{X}$ so that 

\begin{align*}
\int_{(A_n^f)^{i_1} \cap (A_n^g)^{i_2} \cap (A_n^h)^{i_3}} p_{n-1}(x) dx = u_{i_1i_2i_3}^*,  \  \forall i_1, i_2, i_3 \in \{0,1\}.
\end{align*}

\begin{figure}[htb]
  \centering
  \includegraphics[scale=0.5]{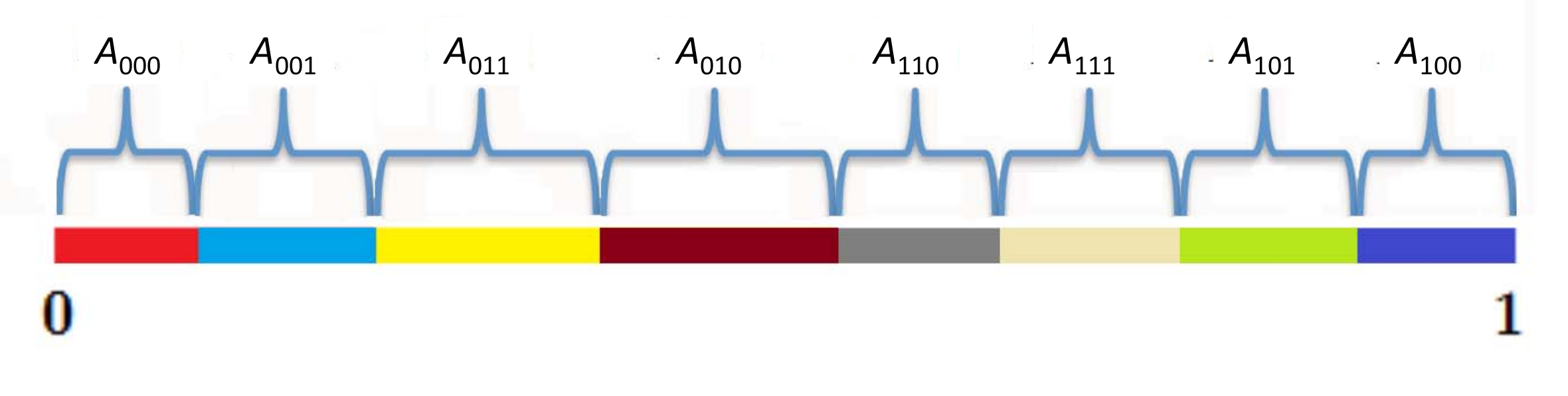}
  \caption{Partition of a line segment into $2^3=8$ disjoint subsets at each stage.}
  \label{fig:linePartition_3sensor}
\end{figure}

To find regions $A_n^f$, $A_n^g$ and $A_n^h$, we first select $\{A_{i_1i_2i_3}$: $i_1, i_2, i_3 \in \{0,1\} \}$ as displayed in Fig. \ref{fig:linePartition_3sensor} such that 
\begin{align*}
\int_{A_{i_1i_2i_3}} p_{n-1}(x) dx = u_{i_1i_2i_3}^*, \  \forall i_1, i_2, i_3 \in \{0,1\}.
\end{align*}

Then, $A_n^f$, $A_n^g$ and $A_n^h$ can be constructed from  $\{A_{i_1i_2i_3}$: $i_1, i_2, i_3 \in \{0,1\} \}$:
\begin{align*}
A_n^f = & \cup_{i_2=0}^1 \cup_{i_3=0}^1 A_{1i_2i_3},  \\
A_n^g = & \cup_{i_1=0}^1 \cup_{i_3=0}^1 A_{i_11i_3},  \\
A_n^h = & \cup_{i_1=0}^1 \cup_{i_2=0}^1 A_{i_1i_21}.
\end{align*}

Similar as in the previous subsection, we conduct 100 Monte Carlo experiments under the optimal joint sensing policy for three Boolean sensors.   In each experiment, the object position $X \in \mathcal{X} = [0,1]$ is randomly generated from a uniform distribution.   The prior density $p_0(x)$ is initialized to be uniform over $\mathcal{X}$, making the initial posterior differential entropy to be zero.   At each stage $n > 0$, we select the sensing modes $A_n^f$, $A_n^g$ and $A_n^h$ based on the current information state $p_{n-1}(x)$ and the optimal joint operating point $\boldsymbol{u}^*$.   The noisy measurements $(y_n^f,y_n^g,y_n^h)$ are randomly generated according to the sensor error models, and will be used to update the conditional density from $p_{n-1}(x)$ to $p_n(x)$.   The process continues until $n=20$ sensing stages are complete.

\begin{figure}[htb]
  \centering
  \includegraphics[scale=0.6]{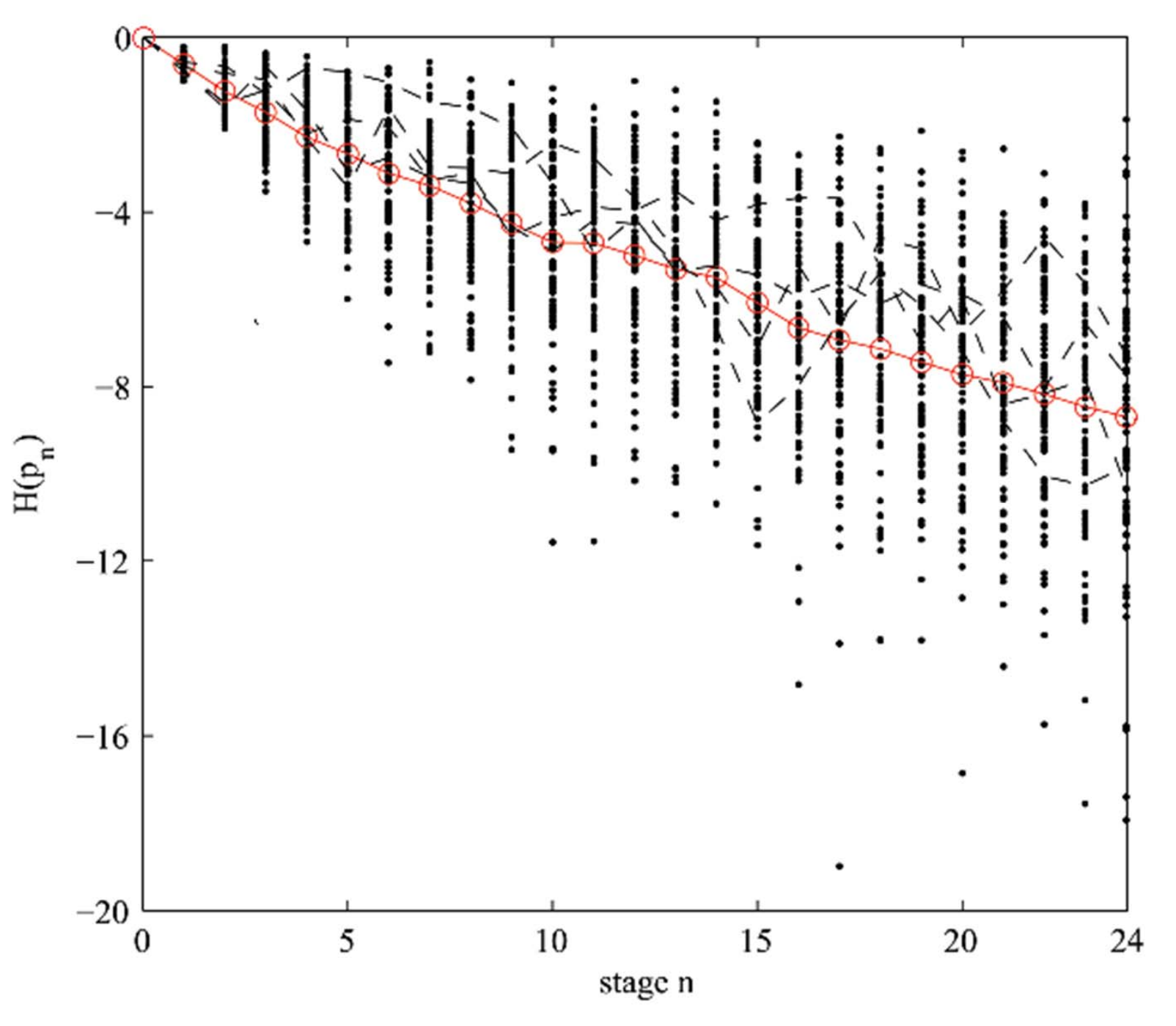}
  \caption{The entropy reduction path for three-sensor joint search under the optimal joint sensing policy.  The red solid line is the average entropy reduction path over 100 realizations, from which we can see that the average entropy reduction decreases linearly.   The specification of the sensors are shown in Table \ref{table:sensorSpecs_3sensor}.   Since the error model of each sensor satisfies the symmetry conditions, the optimal joint operating point is $\boldsymbol{u}^* = (\frac{1}{8}, \frac{1}{8}, \frac{1}{8}, \frac{1}{8}, \frac{1}{8}, \frac{1}{8}, \frac{1}{8}, \frac{1}{8})$.}
  \label{fig:sim_3sensor}
\end{figure}

We compute the differential entropy of the conditional probability density of the object position at each stage of the 100 sample experiments.   The results are plotted in Fig. \ref{fig:sim_3sensor}.   The 100 sample values of $H(p_n)$ are displayed in black dots at each stage $n$.   Four sample trajectories of experiments are shown in dashed lines in order to show the randomness in the actual trajectories.   The average posterior differential entropy reduction path of the 100 sample paths is shown by the red solid line.    It is clear that the average posterior differential entropy decays linearly as predicted by the optimal value function in Proposition \ref{prop:greedyIsOptThm_twosensor}.

\section{Conclusion}
\label{sec:conclusion}

In this paper, we studied the problem of optimal adaptive search for a stationary object under the condition of noisy sensor observations.   
We generalized the formulation of \cite{20qs} in two directions, first by allowing sensors to pose multi-valued queries, and second, by allowing the use of teams of sensors, as in \cite{hero2014collaborative}.  We posed the adaptive sensing problem as a finite horizon stochastic control problem, using a Bayesian formulation for information processing.  The objective function was to minimize the posterior differential entropy of the conditional density of the object location after a finite number of observations.

For the multi-region single-sensor problem, we characterized the optimal sensing strategies as those that maximally reduce the posterior differential entropy at each stage, thus resulting in a myopic or greedy strategy. These optimal sensing strategies must select sensing modes that partition the space into regions that have conditional probabilities of containing the object equal to a fixed vector of probabilities, denoted as the operating point.   We derived an explicit solution for the optimal value function, and showed that such myopic strategies satisfy Bellman's equation of stochastic dynamic programming.  Furthermore,   we provided a convex optimization algorithm for computing the operating point for optimal strategies, and a constructive procedure for computing the optimal sensing actions in real time. 

For the Boolean multi-sensor problems, we considered the case of sensors with general discrete-output error models.  We showed that, as in \cite{hero2014collaborative}, the optimal strategies can be obtained using myopic strategies.  We also established that the joint convex optimization problem for the joint operating point of the optimal strategies can be decoupled into individual scalar convex optimization problems, leading to a simple computational procedure for the solution of multisensor problems.  In addition, we developed sufficient conditions for characterizing symmetry properties of general error models that allow for the analytic solution for the joint operating point of the optimal strategies.  

We extended our multi-sensor formulation to the case sensors can choose between different precision modes as well as sensor modes, at a cost that depends on the precision mode selected.  A choice of precision mode changes the accuracy of the sensor error model.  Our results extend the single sensor results in \cite{microscopy}.  For this case, we develop an explicit solution to the optimal value function for the stochastic control problem, and characterize optimal strategies in terms of selection of a joint operating point and joint precision mode.  We show that this joint sensing mode and precision mode selection can be decoupled into single-sensor scalar convex optimization problems, thereby providing an efficient solution for constructing joint optimal sensing strategies.  

An important note about our results is that they are tied intrinsically to the choice of performance criterion for the stochastic control problem, and the fact that the unknown object location has an absolutely continuous probability distribution with respect to Lebesgue measure.  The choice of posterior Shannon differential entropy as the primary measure of performance allows the application of many concepts from information theory that lead to the explicit optimal solutions derived in this paper.  The fact that the object location distribution has a density allows us to generate sensor modes that can achieve the probabilities determined by the optimal operating points.  Changing the primary objective function to related information measures such as R\'enyi  entropy would invalidate many of the optimality results derived in this paper.

There are several directions in which this paper can be extended.  One important direction is to develop approaches for approximating the optimal search strategies when sensors consist of physical platforms that must move over the region of interest to do the search.  Similar issues arise in the results in classical search theory \cite{StoneBook} which computes optimal allocation of search effort without focusing on individual platforms.  Developing physically realizable sensor plans is necessary for implementation in multiple sensor platforms.  Another extension is to consider problems where platform motion results in constraints as to how sensor modes can change sequentially.  A third extension is to consider problems where sensors have constraints on the types of areas that sensors can observe, based on geometric constraints on sensor field of view.  Other extensions include problems where multiple objects are present and need to be localized, and problems where different sensor modes have mode-dependent costs.  It is unlikely that the structures exploited in these problems will generalize to those formulations, so the focus will be on getting lower bounds on performance, and developing practical approximation strategies based on such bounds.

\vspace{20pt}
{\Large \noindent {\bf Appendix}}
\vspace{20pt}
{

\noindent \bf{Proof of Proposition \ref{prop:OneStepEntropyUpdate_multianswer}}
\begin{proof}
Let $\eta_0(y,x) = \sum_{k=1}^K f_k(y) \mathbbm{1}_{\{x\in A^{(k)}_{n+1}\} } $, $\eta(y) = \int_{x \in  \mathcal{X}} p_n(x) \eta_0(y,x) dx =\sum_{k=1}^K u^{(k)}_{n+1} f_k(y) $.  Then, 
using Bayes' rule as in \eqref{Bayesrule}, we get

\begin{small}
\begin{align*}
 E_{Y_{n+1}} &[H(p_{n+1})|A_{n+1},p_n] 
=  \sum_{y \in \mathcal{Y}}  H(p_{n+1}) P(Y_{n+1}=y|A_{n+1},p_n)  \\
= & - \sum_{y \in \mathcal{Y}} \eta(y) \Big(\int_{\mathcal{X}} [p_n(x) \frac{\eta_0(y,x)}{\eta(y)}] \log[p_n(x)\frac{\eta_0(y,x)}{\eta(y)}]dx\Big)    - \int_{\mathcal{X}} p_n(x)\log{p_n(x)}dx\\
& + \sum_{y \in \mathcal{Y}}\int_{\mathcal{X}}  p_n(x)  \eta_0(y,x) \log{\eta(y)} dx   -   \int_{\mathcal{X}} p_n(x) \sum_{y \in \mathcal{Y}}  \eta_0(y,x) \log{\eta_0(y,x)} dx \\
= & H(p_n) - \bigg[ \mathcal{H}(\eta(y)) - \sum_{k=1}^K \int_{\mathcal{X}} p_n(x) \mathcal{H}(f_k) 
\mathbbm{1}_{\{x \in A^{(k)}_{n+1}\}} dx \bigg]\\
= & H(p_n) - \bigg[ \mathcal{H}(\sum_{k=1}^K u^{(k)}_{n+1} f_k) - \sum_{k=1}^K u^{(k)}_{n+1} \mathcal{H}(f_k) \bigg]
\end{align*}
\end{small}
\end{proof}

\noindent \bf{Proof of Proposition \ref{prop:greedyIsOptThm_multianswer}}
\begin{proof}
To establish this, we show that \eqref{eq:value_fun_multianswer} satisfies the Bellman equation \eqref{BellmanEquation} and the above policy is a minimizing policy.  The optimal value function is is correct at stage $N$, as $V(p_N,N) = H(p_N)$.  Assume by induction that the optimal value function satisfies \eqref{eq:value_fun_multianswer} for all $k \ge n+1$.  Then, 
\vspace{-2pt}
\begin{align*}
V(p_n&,n)
= \inf_{A} E_{Y_{n+1}} [V(p_{n+1},n+1)|A_{n+1}=A,p_n]\\
&= \inf_{A} E_{Y_{n+1}}[H(p_{n+1}) - (N-n-1)\varphi^*)|A_{n+1}=A,p_n]\\
&= \inf_{A} E_{Y_{n+1}}[H(p_{n+1})|A_{n+1}=A,p_n] - (N-n-1)\varphi^*\\
&= H(p_n) -    \sup_{A} \bigg[ \mathcal{H}(\sum_{k=1}^K u^{(k)}_{n+1} f_k) - \sum_{k=1}^K u^{(k)}_{n+1} \mathcal{H}(f_k) \bigg]  - (N-n-1)\varphi^*
\end{align*}

\vspace{-5pt}
\noindent
because of Proposition \ref{prop:OneStepEntropyUpdate_multianswer}.  Furthermore, we know that 
$$ 
\sup_{A} \bigg[ \mathcal{H}(\sum_{k=1}^K u^{(k)}_{n+1} f_k) - \sum_{k=1}^K u^{(k)}_{n+1} \mathcal{H}(f_k) \bigg] = \varphi^*
$$
because, given $p_n$, there is a partition $A$ such that $u^{(k)}(A^{(k)}) = u^{(k)*}$.  Thus, 
$$ V(p_n,n) = V(p_n,n) - \varphi^* -   (N-n-1)\varphi^* =  V(p_n,n) - \varphi^* -  (N-n-)\varphi^*$$
 Furthermore, we have already provided a construction for choosing  $A_{n+1}$ that achieves the supremum:  $\boldsymbol{u}(A_{n+1},p_n) = \boldsymbol{u}^*$.

\end{proof}

\noindent \bf{Proof of Proposition \ref{prop:mse_lower_bound}}
\begin{proof}
Let $\hat X_n = \int_{\mathcal{X}} x p_n(x) dx$ and $\Sigma_n = E[(X - \hat X_n) (X - \hat X_n)^T]$.  By Theorem 17.2.3 in \cite{CoverBook} and Jensen's inequality, under any policy $\zeta$, we have
\begin{align*}
  E_{\zeta} [H(p_n)] 
\leq & E_{\zeta} \Big[\frac{1}{2} \log((2\pi e)^d \det( \Sigma_n))\Big]  \\
\leq & \frac{1}{2} \log(2 \pi e)^d + \frac{1}{2} \log (\det (E_{\zeta} [\Sigma_n]))  \\
= & \frac{1}{2} \log ((2 \pi e)^d \det (E_{\zeta} [\Sigma_n]))
\end{align*}
where $\det(\cdot)$ denotes the matrix determinant.  From Proposition \ref{prop:greedyIsOptThm_multianswer}, under any policy $\zeta$, we  have $E_{\zeta}[H(p_n)] \geq H(p_0) - n\varphi^*$.  By letting $C_0 = e^{2H(p_0)}$, we have
\begin{align*}
\frac{C_0 e^{-2n\varphi^*}}{(2\pi e)^d} \leq \frac{e^{2E_{\zeta}[H(p_n)]}}{(2\pi e)^d}
\leq \det(E_{\zeta}[\Sigma_n])
\end{align*}
where $\mbox{tr}(\cdot)$ denotes the matrix trace.   Since the determinant and the trace of a square matrix can be written as the product and the sum of the eigenvalues of the matrix respectively, using the inequality of arithmetic and geometric means we have
\begin{align*}
\det(E_{\zeta}[\Sigma_n]) \leq \Big( \frac{E_{\zeta} [\mbox{tr}(\Sigma_n)]}{d} \Big)^d 
\end{align*}

Combining and rewriting the inequalities above, we get
\begin{align*}
E[||X - \hat X_n||_2^2] = E_{\zeta} [\mbox{tr}(\Sigma_n)] \geq \frac{d \sqrt[d]{C_0}}{2 \pi e} e^{-\frac{2n\varphi^*}{d}} 
\end{align*}
\end{proof}

\noindent \bf{Proof of Proposition \ref{prop:OneStepEntropyUpdate_twosensor}}
\begin{proof}

Let $\eta_0(\boldsymbol{y},x)$ be defined as 
$$\eta_0(\boldsymbol{y},x) = P(\boldsymbol{Y}_{n+1} = \boldsymbol{y} | \boldsymbol{A}_{n+1},X = x)$$
Then,
\begin{small}
\begin{align*}
  E_{\boldsymbol{Y}_{n+1}}& [H(p_{n+1}) | \boldsymbol{A}_{n+1} = (A^{(1)}, \ldots, A^{(M)}), p_n] \\
=& \sum_{\boldsymbol{y} \in \mathcal{Y}^M} H(p_{n+1}) P(\boldsymbol{Y}_{n+1} = \boldsymbol{y} = (y^{(1)},\ldots,y^{(M)}) | \boldsymbol{A}_{n+1} = (A^{(1)}, \ldots, A^{(M)}),p_n)   \\
= & - \sum_{\boldsymbol{y} \in \mathcal{Y}^M} \eta(\boldsymbol{y}) \Big( \int_{\mathcal{X}} [p_n(x)\frac{\eta_0(\boldsymbol{y},x)}{\eta(\boldsymbol{y})}] \log[p_n(x)\frac{\eta_0(\boldsymbol{y},x)}{\eta(\boldsymbol{y})}] dx \Big)   \\
= & - \int_{\mathcal{X}} p_n(x) \log p_n(x) dx + \sum_{\boldsymbol{y} \in \mathcal{Y}^M} \int_{\mathcal{X}} p_n(x) \eta_0(\boldsymbol{y},x) \log \eta(\boldsymbol{y}) dx   - \int_{\mathcal{X}} p_n(x) \sum_{\boldsymbol{y} \in \mathcal{Y}^M} \eta_0(\boldsymbol{y},x) \log \eta_0(\boldsymbol{y},x) dx  \\
= & H(p_n) - \Big[ \mathcal{H}(\eta(\boldsymbol{y})) - \sum_{i_{1}=0}^1\cdots \sum_{i_M=0}^1 \int_{\mathcal{X}} p_n(x) \mathcal{H}(q_{i_{1:M}}(\boldsymbol{y})) \mathbbm{1}_{\{x \in \cap_{m=1}^M (A^{(m)})^{i_m} \}} dx \Big]   \\
= & H(p_n) - \Big[ \mathcal{H}(\sum_{i_{1}=0}^1\cdots \sum_{i_M=0}^1 q_{i_{1:M}}(\boldsymbol{y}) u_{i_{1:M}}) - \sum_{i_{1}=0}^1\cdots \sum_{i_M=0}^1 u_{i_{1:M}} \mathcal{H}(q_{i_{1:M}}) \Big]
\end{align*}
\end{small}
\end{proof}

\noindent \bf{Proof of Proposition \ref{prop:multi_realizable}}
\begin{proof}
The proposition follows because $X \in \mathbb{R}^d$ is a continuous random variable with continuous cumulative probability distribution, whose posterior density at stage $n$ is $p_{n-1}(x)$.  Thus, given $\boldsymbol{u}^* \ge 0$, $\sum_{i_{1}=0}^1\cdots \sum_{i_M=0}^1 u_{i_{1:M}}^* = 1$, we can use our results for the multiregion single sensor problem to find a  partition of the domain of $X$ into $2^M$ disjoint subsets $\{B_{i_{1:M}}: i_{1:M} \in \{0,1\}^M \}$ such that the probability of each subset is $\int_{B_{i_{1:M}}} p_{n-1}(x) dx = u_{i_{1:M}}^*$, $\forall i_{1:M} \in \{0,1\}^M $.  Then by letting $A^{(m)} = \cup_{\{i_{1:M} : i_m=1\}} B_{i_{1:M}}$, $\forall m=1,\ldots,M$, we can realize $\boldsymbol{u}^*$.
\end{proof}

\noindent \bf{Proof of Proposition \ref{thm:equiv}}
\begin{proof}
We first prove that $\varphi^*  \leq  \sum_{m=1}^M \varphi^{(m)*}$: 
\begin{small}
\begin{align*}
  \varphi^*  
= & \mathcal{H}(\sum_{i_{1}=0}^1 \cdots \sum_{i_{M}=0}^1  u_{i_{1:M}}^* \cdot q_{i_{1:M}})  -  \sum_{i_{1}=0}^1 \cdots \sum_{i_{M}=0}^1 u_{i_{1:M}}^* \mathcal{H}(q_{i_{1:M}})  
\end{align*}
\end{small}
From the additivity property of the Shannon entropy, we have:
$$ \mathcal{H}(q_{i_{1:M}}) = \sum_{m=1}^M (\mathcal{H}(f_1^{(m)})  \mathbbm{1}_{\{i_m = 1\}} + \mathcal{H}(f_0^{(m)})  \mathbbm{1}_{\{i_m = 0\}} ) $$
Thus, 
 $$\sum_{i_{1}=0}^1 \cdots \sum_{i_{M}=0}^1 u_{i_{1:M}}^* \mathcal{H}(q_{i_{1:M}}) = \sum_{m=1}^M \Bigl( (\sum_{i_{1:M}: i_m=1} u_{i_{1:M}}^*) \mathcal{H} (f_1^{(m)})+
(\sum_{i_{1:M}: i_m=0} u_{i_{1:M}}^*) \mathcal{H} (f_0^{(m)}) \Bigr) $$
Similarly, note that the term $\sum_{i_{1}=0}^1 \cdots \sum_{i_{M}=0}^1  u_{i_{1:M}}^* \cdot q_{i_{1:M}}$ specifies a joint probability distribution for the variables $Y^{(1)}, \ldots, Y^{(m)}$, with marginal probability distribution for each variable $Y^{(m)}$ given by 
$$g^{(m)}(y) = (\sum_{i_{1:M}: i_m=1} u_{i_{1:M}}^*) f_1^{(m)}(y) + (\sum_{i_{1:M}: i_m=0} u_{i_{1:M}}^*) f_0^{(m)}(y) $$
Combining these relations and using the subadditivity property of the Shannon entropy, we obtain
\begin{small}
\begin{align*}
  \varphi^*  
\leq &  \sum_{m=1}^M \mathcal{H}(g^{(m)}) -\sum_{m=1}^M \Bigl( (\sum_{i_{1:M}: i_m=1} u_{i_{1:M}}^*) \mathcal{H} (f_1^{(m)})+
(\sum_{i_{1:M}: i_m=0} u_{i_{1:M}}^*) \mathcal{H} (f_0^{(m)})   \Bigr)
\end{align*}
\end{small}
Note that the numbers $a^{(m)} = \sum_{i_{1:M}: i_m=1} u_{i_{1:M}}^*$ and $b^{(m)} =  \sum_{i_{1:M}: i_m=0} u_{i_{1:M}}^*$ are non-negative and sum up to 1, and thus represent a possible operating point for sensor $m$.  
Since $(u^{(m)*}, 1-u^{(m)*})$  is the optimal operating point that maximizes $\varphi^{(m)}(u)$, we have 
\begin{small}
\begin{align*}
  \varphi^*  
\leq &  \sum_{m=1}^M \mathcal{H}(g^{(m)}) -\sum_{m=1}^M \Bigl( (\sum_{i_{1:M}: i_m=1} u_{i_{1:M}}^*) \mathcal{H} (f_1^{(m)})+
(\sum_{i_{1:M}: i_m=0} u_{i_{1:M}}^*) \mathcal{H} (f_0^{(m)}) \Bigr)  \\
=& \sum_{m=1}^M \varphi(a^{(m)}) \leq \sum_{m=1}^M \varphi^{(m)*}  
\end{align*}
\end{small}

Given ${u^{(m)*}}$, define
\begin{small}
\begin{align*}
\overline{u}_{i_{1:M}} = \prod_{m=1}^M (u^{(m)*})^{i_m} (1-u^{(m)*})^{1-i_m} 
\end{align*}
\end{small}
\noindent
Note that $\sum_{i_{1}=0}^1\cdots \sum_{i_M=0}^1 \overline{u}_{i_{1:M}} =1 $, so this is a valid joint operating point $\boldsymbol{u}$ for the multisensor problem.  Then, 
\begin{small}
\begin{align*}
  \varphi&(\boldsymbol{\overline{u}}) 
=  \mathcal{H}(\sum_{i_{1}=0}^1\cdots \sum_{i_M=0}^1  \overline{u}_{i_{1:M}} \cdot q_{i_{1:M}})  -  \sum_{i_{1}=0}^1\cdots \sum_{i_M=0}^1 \overline{u}_{i_{1:M}} \mathcal{H}(q_{i_{1:M}}) \nonumber\\
= & \mathcal{H}(\sum_{i_{1}=0}^1\cdots \sum_{i_M=0}^1  \prod_{m=1}^M (u^{(m)*})^{i_m} (1-u^{(m)*})^{1-i_m} q_{i_{1:M}})  -  \sum_{i_{1}=0}^1\cdots \sum_{i_M=0}^1 (\prod_{m=1}^M (u^{(m)*})^{i_m} (1-u^{(m)*})^{1-i_m}) \mathcal{H}(q_{i_{1:M}}) \nonumber\\
= & \mathcal{H} \Big( \prod_{m=1}^M [\sum_{j=0}^1 f_j^{(m)} (u^{(m)*})^j (1-u^{(m)*})^{(1-j)}] \Big) - \sum_{m=1}^M \sum_{j=0}^1 (u^{(m)*})^j (1-u^{(m)*})^{(1-j)} \mathcal{H} (f_j^{(m)})   \nonumber\\
= & \sum_{m=1}^M  \Big[  \mathcal{H} \Big( \sum_{j=0}^1 f_j^{(m)} (u^{(m)*})^j (1-u^{(m)*})^{(1-j)} \Big) - \sum_{j=0}^1 (u^{(m)*})^j (1-u^{(m)*})^{(1-j)} \mathcal{H} (f_j^{(m)})  \Big]   \nonumber\\
= & \sum_{m=1}^M  \varphi^{(m)*}
\end{align*}
\end{small}

\vspace{-10pt}
Since $\varphi^* = \max_{\boldsymbol{u}} \varphi(\boldsymbol{u}) \leq \sum_{m=1}^M  \varphi^{(m)*}$ and $\varphi(\boldsymbol{u})$ has a unique optimal point, selecting $\boldsymbol{u}$ as \eqref{eq:uopt_from_ufug} will give us the optimal operating point for $\varphi(\boldsymbol{u})$ and we have $\varphi^* = \sum_{m=1}^M  \varphi^{(m)*}$.
\end{proof}

\noindent \bf{Proof of Lemma \ref{lemma:oneStage_precisionMode}}
\begin{proof}

\begin{small}
\begin{align*}
  \hat G (p_n,& \boldsymbol{A}, \boldsymbol{l})  \\
= & H(p_n) - E_{\boldsymbol{Y}_{n+1}} [H(p_{n+1}) + \gamma \sum_{m=1}^M W^{(m)}(l^{(m)}) | (\boldsymbol{A}, \boldsymbol{l}), p_n)]  \\
= & H(p_n) - E_{\boldsymbol{Y}_{n+1}} [H(p_{n+1}) | (\boldsymbol{A}, \boldsymbol{l}), p_n] - \gamma \sum_{m=1}^M W^{(m)}(l^{(m)})   \\
= & H(p_n) - \Big( \mathcal{H}(p_n) - \Big[ \mathcal{H}(\sum_{i_{1}=0}^1\cdots \sum_{i_M=0}^1 q_{i_{1:M}}^{\boldsymbol{l}} u_{i_{1:M}}) - \sum_{i_{1}=0}^1\cdots \sum_{i_M=0}^1 u_{i_{1:M}} \mathcal{H}(q_{i_{1:M}}^{\boldsymbol{l}}) \Big] \Big) - \gamma \sum_{m=1}^M W^{(m)}(l^{(m)})    \\
= & \Big[ \mathcal{H}(\sum_{i_{1}=0}^1\cdots \sum_{i_M=0}^1 q_{i_{1:M}}^{\boldsymbol{l}} u_{i_{1:M}}) - \sum_{i_{1}=0}^1\cdots \sum_{i_M=0}^1 u_{i_{1:M}} \mathcal{H}(q_{i_{1:M}}^{\boldsymbol{l}}) \Big] - \gamma \sum_{m=1}^M W^{(m)}(l^{(m)})   \\
= & G(\boldsymbol{u}, \boldsymbol{l})
\end{align*}
\end{small}

\end{proof}

\noindent \bf{Proof of Proposition \ref{prop:optimal_precision_mode}}
\begin{proof}
The optimal value function $V(p_N,N) = {H}(p_N) - (N-N) G^* = H(p_N)$ satisfies the hypothesized form at the terminal time $N$.  We show by induction that the optimal value function has the postulated form, and that the optimal strategies in the theorem achieve the infimum in Bellman's equation:
\begin{align*}
V(p_n, n) = \inf_{\boldsymbol{A}_{n+1}, \boldsymbol{l}_{n+1}} E_{\mathbf{Y}_{n+1}}[V(p_{n+1}, n+1) + \gamma \sum_{m=1}^M W^{(m)}(l_{n+1}^{(m)})  | (\boldsymbol{A}_{n+1}, \boldsymbol{l}_{n+1}), p_n]
\end{align*}
Assuming that $V(p_{n+1},n+1) = H(p_{n+1}) + (N - n - 1) G^*$, we have 
\begin{small}
\begin{align*}
E_{\mathbf{Y}_{n+1}}&
[V(p_{n+1}, n+1) + \gamma \sum_{m=1}^M W^{(m)}(l_{n+1}^{(m)})  | (\boldsymbol{A}_{n+1}, \boldsymbol{l}_{n+1}), p_n]\\&= E_{\mathbf{Y}_{n+1}}[H(p_{n+1}) + (N-n-1)G^* + \gamma \sum_{m=1}^M W^{(m)}(l_{n+1}^{(m)})  | (\boldsymbol{A}_{n+1}, \boldsymbol{l}_{n+1}), p_n]\\
&= H(p_n) - G(\boldsymbol{u}_{n+1},\boldsymbol{l}_{n+1}) + (N - n - 1) G^*
\end{align*}
\end{small}
by Lemma \ref{lemma:oneStage_precisionMode}.

For fixed $\boldsymbol{l}$, $G(\boldsymbol{u}, \boldsymbol{l})$ is strictly concave over $\sum_{i_{1}=0}^1\cdots \sum_{i_M=0}^1 u_{i_{1:M}} = 1$ due to the strict concavity of the Shannon entropy.   Thus, 
$$G^* = \inf_{\boldsymbol{u},\boldsymbol{l}} G(\boldsymbol{u},\boldsymbol{l})$$
Furthemore, we know from the discussion after Proposition \ref{prop:OneStepEntropyUpdate_multianswer} and Proposition \ref{prop:multi_realizable} that, for any density $p_n(x)$ and desired joint operating point $\boldsymbol{u}$, there exists a joint sensing mode $\boldsymbol{A}_{n+1}$ such that the probabilities $\boldsymbol{u}(\boldsymbol{A}_{n+1}, p_{n+1}) = \boldsymbol{u}$.
Thus, 
\begin{small}
\begin{align*}
V(p_n, n) &= \inf_{\boldsymbol{A}_{n+1}, \boldsymbol{l}_{n+1}} [H(p_n) - G(\boldsymbol{u}_{n+1},\boldsymbol{l}_{n+1}) + (N - n - 1) G^*]\\
&= H(p_n) - \sup_{\boldsymbol{A}_{n+1}, \boldsymbol{l}_{n+1}}G(\boldsymbol{u}_{n+1},\boldsymbol{l}_{n+1}) - (N - n - 1)G^* \\
&= H(p_n) - (N-n) G^*
\end{align*}
\end{small}

Note that 
$$\sup_{\boldsymbol{u},\boldsymbol{l}} G(\boldsymbol{u},\boldsymbol{l}) = \max_{\boldsymbol{u},\boldsymbol{l}} G(\boldsymbol{u},\boldsymbol{l})$$
is achieved at some $(\boldsymbol{u}^*,\boldsymbol{l}^*)$ because it is the maximum of a finite number of strictly concave functions defined over the compact $M$-dimensional simplex.  Thus, the optimal strategies are given by
any $(\boldsymbol{A}_{n+1}, \boldsymbol{l}_{n+1})$ such that $\boldsymbol{l}_{n+1} = \boldsymbol{l}^*$, and $ \boldsymbol{u}(\boldsymbol{A}_{n+1}, p_{n+1}) = \boldsymbol{u}^*$.

\end{proof}

\noindent {\bf Proof of Proposition \ref{thm:equiv_prec}}
\begin{proof}
We first prove that $G^*  \leq  \sum_{m=1}^M G^{(m)*}$: 
\begin{small}
\begin{align*}
  G^*   
= & \mathcal{H}(\sum_{i_{1}=0}^1\cdots \sum_{i_{M}=0}^1  u_{i_{1:M}}^* \cdot q_{i_{1:M}}^{\boldsymbol{l}^*})  - \sum_{i_{1}=0}^1\cdots \sum_{i_{M}=0}^1 u_{i_{1:M}}^* \mathcal{H}(q_{i_{1:M}}^{\boldsymbol{l}^*})    -    \gamma \sum_{m=1}^M W^{(m)} (l^{(m)*})    \\
\leq & \sum_{m=1}^M \Big[ 
\mathcal{H} \Big( (\sum_{i_{1:M}: i_m=1} u_{i_{1:M}}^*) f_1^{(m, l^{(m)*})} + (\sum_{i_{1:M}: i_m=0} u_{i_{1:M}}^*) f_0^{(m, l^{(m)*})} \Big)   \nonumber\\
& - \Big( (\sum_{i_{1:M}: i_m=1} u_{i_{1:M}}^*) \mathcal{H} (f_1^{(m, l^{(m)*})}) 
+ (\sum_{i_{1:M}: i_m=0} u_{i_{1:M}}^*) \mathcal{H} (f_0^{(m, l^{(m)*})}) \Big)     - \gamma W^{(m)} (l^{(m)*})
\Big]   \\
\leq & \sum_{m=1}^M \Big[
\mathcal{H} \Big( u^{(m)*} f_1^{(m, l^{(m)*})} + (1-u^{(m)*}) f_0^{(m, l^{(m)*})} \Big)   \\
& - \Big( u^{(m)*} \mathcal{H} (f_1^{(m, l^{(m)*})}) 
+ (1-u^{(m)*}) \mathcal{H} (f_0^{(m, l^{(m)*})}) \Big)  - \gamma  W^{(m)} (l^{(m)*})  
\Big]   \\
= & \sum_{m=1}^M G^{(m)*}     
\end{align*}
\end{small}

The first inequality results from the subadditivity and additivity properties of the Shannon entropy. The second inequality is true because $(u^{(m)*}, l^{(m)*})$, $m=1,\ldots,M$, are the optimal points of $G^{(m)}(u^{(m)}, l^{(m)})$, $m=1,\ldots,M$, respectively.

Let
\begin{small}
\begin{align*}
\overline{u}_{i_{1:M}} = \prod_{m=1}^M (u^{(m)*})^{i_m} (1-u^{(m)*})^{1-i_m}; \quad \overline{\boldsymbol{l}} = (l^{(1)*},\ldots,l^{(M)*}) 
\end{align*}
\end{small}

\vspace{-7pt}
\noindent
Plug them into $G(\boldsymbol{u}, \boldsymbol{l})$, we have
\begin{small}
\begin{align*}
   G(\boldsymbol{\overline{u}}, \boldsymbol{\overline{l}})  
= & \mathcal{H}(\sum_{i_{1}=0}^1\cdots \sum_{i_{M}=0}^1  \overline{u}_{i_{1:M}} \cdot q_{i_{1:M}}^{\boldsymbol{\overline{l}}})  -  \sum_{i_{1}=0}^1\cdots \sum_{i_{M}=0}^1 \overline{u}_{i_{1:M}} \mathcal{H}(q_{i_{1:M}}^{\boldsymbol{\overline{l}}})  
- \gamma \sum_{m=1}^M W^{(m)}(\overline{l}^{(m)})      \\
= & \mathcal{H} \Big( \prod_{m=1}^M [\sum_{j=0}^1 f_j^{(m, l^{(m)*})} (u^{(m)*})^j (1-u^{(m)*})^{(1-j)}] \Big)     \\
  & - \sum_{m=1}^M \sum_{j=0}^1 (u^{(m)*})^j (1-u^{(m)*})^{(1-j)} \mathcal{H} (f_j^{(m, l^{(m)*})})  - \gamma \sum_{m=1}^M W^{(m)}(l^{(m)*})      \\
= & \sum_{m=1}^M  \Big[  \mathcal{H} \Big( \sum_{j=0}^1 f_j^{(m, l^{(m)*})} (u^{(m)*})^j (1-u^{(m)*})^{(1-j)} \Big)    \\
  & - \sum_{j=0}^1 (u^{(m)*})^j (1-u^{(m)*})^{(1-j)} \mathcal{H} (f_j^{(m, l^{(m)*})})  -  \gamma W^{(m)}(l^{(m)*}) \Big]   \nonumber\\
= & \sum_{m=1}^M  G^{(m)*}
\end{align*}
\end{small}

\vspace{-10pt}
Since $G^* = \max_{(\boldsymbol{u}, \boldsymbol{l})} G(\boldsymbol{u}, \boldsymbol{l}) \leq \sum_{m=1}^M  G^{(m)*}$, selecting $(\boldsymbol{u}, \boldsymbol{l})$ as \eqref{eq:full_from_sing} will give us an optimal operating point for $G(\boldsymbol{u}, \boldsymbol{l})$ and we have $G^* = \sum_{m=1}^M  G^{(m)*}$.
\end{proof}

}

\bibliographystyle{IEEETran}  
\bibliography{references}

\begin{thebibliography}{10}
\providecommand{\url}[1]{#1}
\csname url@samestyle\endcsname
\providecommand{\newblock}{\relax}
\providecommand{\bibinfo}[2]{#2}
\providecommand{\BIBentrySTDinterwordspacing}{\spaceskip=0pt\relax}
\providecommand{\BIBentryALTinterwordstretchfactor}{4}
\providecommand{\BIBentryALTinterwordspacing}{\spaceskip=\fontdimen2\font plus
\BIBentryALTinterwordstretchfactor\fontdimen3\font minus
  \fontdimen4\font\relax}
\providecommand{\BIBforeignlanguage}[2]{{%
\expandafter\ifx\csname l@#1\endcsname\relax
\typeout{** WARNING: IEEEtran.bst: No hyphenation pattern has been}%
\typeout{** loaded for the language `#1'. Using the pattern for}%
\typeout{** the default language instead.}%
\else
\language=\csname l@#1\endcsname
\fi
#2}}
\providecommand{\BIBdecl}{\relax}
\BIBdecl

\bibitem{Koopman1946}
B.~O. Koopman, ``{Search and Screening. Operations Evaluation Group Report No.\
  56},'' Center for Naval Analyses, Alexandria, VA, Tech. Rep., 1946.

\bibitem{KoopmanBook}
------, \emph{Search and Screening: General Principles with Historical
  Applications}.\hskip 1em plus 0.5em minus 0.4em\relax Pergamon, New York NY,
  1980.

\bibitem{StoneBook}
L.~D. Stone, \emph{Theory of optimal search}.\hskip 1em plus 0.5em minus
  0.4em\relax Academic Press New York, 1975.

\bibitem{Stone1977}
------, ``Search theory: a mathematical theory for finding lost objects,''
  \emph{Mathematics Magazine}, pp. 248--256, 1977.

\bibitem{Benkoski1991}
S.~J. Benkoski, M.~G. Monticino, and J.~R. Weisinger, ``A survey of the search
  theory literature,'' \emph{Naval Research Logistics}, vol.~38, no.~4, pp.
  469--494, 1991.

\bibitem{Castanon1995}
D.~A. Casta{\~n\'o}n, ``Optimal search strategies in dynamic hypothesis
  testing,'' \emph{IEEE Transactions on Systems, Man and Cybernetics}, vol.~25,
  no.~7, pp. 1130--1138, Jul 1995.

\bibitem{DavidsBook}
A.~O. Hero, D.~A. Casta{\~n\'o}n, D.~Cochran, and K.~Kastella,
  \emph{Foundations and Applications of Sensor Management}.\hskip 1em plus
  0.5em minus 0.4em\relax Springer, 2008.

\bibitem{acLearn1}
R.~Castro and R.~Nowak, ``Active learning and sampling,'' in \emph{Foundations
  and Applications of Sensor Management}, A.~Hero, D.~A. Casta{\~n}{\'o}n,
  D.~Cochran, and K.~Kastella, Eds.\hskip 1em plus 0.5em minus 0.4em\relax
  Springer, 2008.

\bibitem{faceDetect}
R.~Sznitman and B.~Jedynak, ``Active testing for face detection and
  localization,'' \emph{IEEE Transactions on Pattern Analysis and Machine
  Intelligence}, 2010.

\bibitem{castanon2005stochastic}
D.~A. Casta{\~n\'o}n, ``Stochastic control bounds on sensor network
  performance,'' in \emph{Proc. IEEE Conference on Decision and Control and
  European Control Conference}, Seville, Spain, 2005, pp. 4939--4944.

\bibitem{recedeHorizon}
D.~C. Hitchings and D.~A. Casta{\~n}{\'o}n, ``Receding horizon stochastic
  control algorithms for sensor management,'' \emph{Proc. American Control
  Conference}, June 2010.

\bibitem{infoBased}
K.~L. Jenkins and D.~A. Casta{\~n}{\'o}n, ``Information-based adaptive sensor
  management for sensor networks,'' \emph{Proc. American Control Conference},
  June 2011.

\bibitem{markovObj}
D.~C. Hitchings and D.~A. Casta{\~n}{\'o}n, ``Sensor control for search and
  identification of markov objects,'' \emph{Proc. IEEE Conference on Decision
  and Control and European Control Conference}, December 2011.

\bibitem{williams_fisher_willsky07}
J.~L. Williams, J.~W. {Fisher III}, and A.~S. Willsky, ``Approximate dynamic
  programming for communication-constrained sensor network management,''
  \emph{IEEE Transactions on Signal Processing}, 2007.

\bibitem{kreucher_hero_kastella_morelande07}
C.~Kreucher, A.~O. {Hero III}, K.~Kastella, and M.~R. Morelande, ``An
  information-based approach to sensor management in large dynamic networks,''
  \emph{Proceedings of IEEE}, 2007.

\bibitem{kreucher2005sensor}
C.~Kreucher, K.~Kastella, and A.~O. Hero~III, ``Sensor management using an
  active sensing approach,'' \emph{Signal Processing}, vol.~85, no.~3, pp.
  607--624, 2005.

\bibitem{dpbook}
D.~P. Bertsekas, \emph{Dynamic Programming and Optimal Control}, 3rd~ed.\hskip
  1em plus 0.5em minus 0.4em\relax Athena Scientific, 2005, vol.~1.

\bibitem{seqExp1}
M.~H. DeGroot, \emph{Optimal Statistical Decisions}.\hskip 1em plus 0.5em minus
  0.4em\relax McGraw Hill, 1970.

\bibitem{seqExp2}
G.~B. Wetherill and K.~D. Glazebrook, \emph{Sequential Methods in Statistics},
  3rd~ed., ser. Monographs on Statistics and Applied Probability.\hskip 1em
  plus 0.5em minus 0.4em\relax Chapman \& Hall, 1986.

\bibitem{seqExp3}
H.~Robbins, ``Some aspects of the sequential design of experiments,''
  \emph{Bulletin of the American Mathematical Society}, 1952.

\bibitem{20qs}
B.~Jedynak, P.~I. Frazier, and R.~Sznitman, ``Twenty questions with noise:
  Bayes optimal policies for entropy loss,'' \emph{Journal of Applied
  Probability}, vol.~49, pp. 114--136, 2011.

\bibitem{CoverBook}
T.~M. Cover and J.~A. Thomas, \emph{Elements of information theory}.\hskip 1em
  plus 0.5em minus 0.4em\relax John Wiley \& Sons, 2012.

\bibitem{horstein1963sequential}
M.~Horstein, ``Sequential transmission using noiseless feedback,'' \emph{IEEE
  Transactions on Information Theory}, vol.~9, no.~3, pp. 136--143, 1963.

\bibitem{BZ_Algorithm}
M.~V. Burnashev and K.~Zigangirov, ``An interval estimation problem for
  controlled observations,'' \emph{Problemy Peredachi Informatsii}, vol.~10,
  no.~3, pp. 51--61, 1974.

\bibitem{nowak2008generalized}
R.~Nowak, ``Generalized binary search,'' in \emph{Proc. Allerton Conference
  Communication, Control, and Computing}, Monticello, IL, 2008, pp. 568--574.

\bibitem{dhagat1992playing}
A.~Dhagat, P.~G{\'a}cs, and P.~Winkler, ``On playing ``twenty questions'' with
  a liar,'' in \emph{Proc. third annual ACM-SIAM symposium on Discrete
  algorithms}, 1992, pp. 16--22.

\bibitem{spencer1992ulam}
J.~Spencer, ``Ulam's searching game with a fixed number of lies,''
  \emph{Theoretical Computer Science}, vol.~95, no.~2, pp. 307--321, 1992.

\bibitem{microscopy}
R.~Sznitman, A.~Lucchi, P.~I. Frazier, B.~Jedynak, and P.~Fua, ``An optimal
  policy for target localization with application to electron microscopy,''
  \emph{Proc. International Conference on Machine Learning}, 2013.

\bibitem{hero2014collaborative}
T.~Tsiligkaridis, B.~M. Sadler, and A.~O. Hero~III, ``Collaborative 20
  questions for target localization,'' \emph{IEEE Transactions on Information
  Theory}, vol.~60, no.~4, pp. 2233--2252, 2014.

\bibitem{hero2015decentralized}
------, ``On decentralized estimation with active queries,'' \emph{IEEE
  Transactions on Signal Processing}, vol.~63, no.~10, pp. 2610--2622, 2015.

\bibitem{DingCastanon15}
H.~Ding and D.~A. Casta{\~n\'on}, ``Optimal solutions for classes of adaptive
  search problems,'' in \emph{Proc. IEEE Conference on Decison and Control},
  Osaka, Japan, December 2015.

\bibitem{Ber_Shreve}
D.~P. Bertsekas and S.~E. Shreve, \emph{{Stochastic Optimal Control: the
  Discrete Time Case}}.\hskip 1em plus 0.5em minus 0.4em\relax Orlando:
  Academic Press, 1978.

\bibitem{Chen00}
P.-N. Chen and F.~Alajaji, \emph{Lectures Notes in Information Theory}, 2000.

\bibitem{Gallager}
R.~G. Gallager, \emph{{Information Theory and Reliable Communication}}.\hskip
  1em plus 0.5em minus 0.4em\relax New York: John Wiley, 1968.

\bibitem{asymcap}
P.-O. Amblard, O.~J. Michel, and S.~Morfu, ``Revisiting the asymmetric binary
  channel: joint noise-enhanced detection and information transmission through
  threshold devices,'' in \emph{Proc. SPIE 5845 Noise in Complex Systems and
  Stochastic Dynamics III}, May 2005.

\end{thebibliography}

\end{document}